%% file: DesignerlyLoop.tex
\documentclass[sigconf,screen]{acmart}

\usepackage{natbib}
\usepackage{xcolor}  
\usepackage[normalem]{ulem}  
\usepackage{soul}       
\usepackage[dvipsnames]{xcolor}

\renewcommand\hl[1]{#1} 

\AtBeginDocument{%
  }

\setcopyright{acmlicensed}
\copyrightyear{2018}
\acmYear{2018}
\acmDOI{XXXXXXX.XXXXXXX}
\acmConference[Conference acronym 'XX]{Make sure to enter the correct
  conference title from your rights confirmation email}{June 03--05,
  2018}{Woodstock, NY}

\acmISBN{978-1-4503-XXXX-X/2018/06}

\newcommand{\DL}{\textit{DesignerlyLoop}}

\newcommand{\panelB}{\textit{Design Intent Canvas}}
\newcommand{\panelC}{\textit{LLM Reasoning Canvas}}

\definecolor{tabblue}{RGB}{31,119,180}
\definecolor{tabyellow}{RGB}{255,219,0}
\definecolor{tabgreen}{RGB}{140,220,100}

\usepackage[utf8]{inputenc}
\usepackage{xcolor}
\usepackage{listings}
\usepackage{tcolorbox}
\usepackage{threeparttable}
\usepackage{pifont}
\usepackage{subcaption}
\newcommand{\cmark}{\ding{51}}%
\definecolor{myyellow}{RGB}{255,242,204}
\definecolor{lightyellow}{RGB}{255,250,230}
\definecolor{codegray}{RGB}{248,248,248}
\definecolor{userrole}{RGB}{207,72,72}
\definecolor{assistantrole}{RGB}{160,32,240}
\definecolor{commentpurple}{RGB}{138,43,226} 
\definecolor{codered}{RGB}{255,0,0} 
\definecolor{commentgreen}{RGB}{0,150,0} 
\definecolor{codebg}{RGB}{245,245,220}  

\lstdefinelanguage{p5js}{
  keywords={let, const, function, return, if, else, for},
  keywordstyle=\color{blue}\bfseries,
  ndkeywords={true,false,null},
  ndkeywordstyle=\color{magenta},
  identifierstyle=\color{black},
  sensitive=true,
  comment=[l]{//},
  commentstyle=\color{commentpurple}\ttfamily,
  stringstyle=\color{codered},
  morestring=[b]',
  morestring=[b"]  
}

\newtcolorbox{appendixbox}{
  colback=gray!5,
  colframe=white,
  boxrule=0pt,
  arc=4pt,
  left=6pt, right=6pt, top=6pt, bottom=6pt,
  fontupper=\small\ttfamily,
}

\begin{document}

\title[\textit{DesignerlyLoop}]{DesignerlyLoop: Forming Design Intent through Curated Reasoning for Human-LLM Alignment}

\renewcommand{\shortauthors}{Trovato et al.}

\author{Anqi Wang}
\email{awangan@connect.ust.hk}
\orcid{0000-0003-4238-6581}
\affiliation{%
  \institution{Hong Kong University of Science and Technology}
  \city{Hong Kong SAR}
  \state{}
  \country{China}
}

\author{Zhengyi Li}
\email{7802220124@csu.edu.cn}
\affiliation{%
  \institution{Central South University}
  \city{Changsha}
  \state{Hunan}
  \country{China}
}

\author{Xin Tong}
\email{xint@hkust-gz.edu.cn}
\orcid{0000-0002-8037-6301}
\affiliation{%
  \institution{Hong Kong University of Science and Technology (Guangzhou)}
  \city{Guangzhou}
  \country{China}}

\author{Pan Hui}
\authornote{Corresponding author}
\orcid{0000-0001-6026-1083}
\affiliation{%
  \institution{Hong Kong University of Science and Technology (Guangzhou)}
  \city{Guangzhou}
  \country{China}
  \email{panhui@hkust-gz.edu.cn}
}
\affiliation{
  \institution{Hong Kong University of Science and Technology}
  \city{Hong Kong SAR}
  \country{China}
  \email{panhui@ust.hk}
}

\begin{abstract}
\input{Sections/0_Abstract}
\end{abstract}

\ccsdesc[500]{Human-centered computing~Human computer interaction (HCI)}
\ccsdesc[500]{Human-centered computing~Empirical studies in HCI}
\ccsdesc[300]{Applied computing~Computer-aided design}
\keywords{creativity support, human-AI collaboration, LLM reasoning, AI-assisted design, Large language model (LLM), LLM chain}

\begin{teaserfigure}
    \centering
    \includegraphics[width=0.8\linewidth]{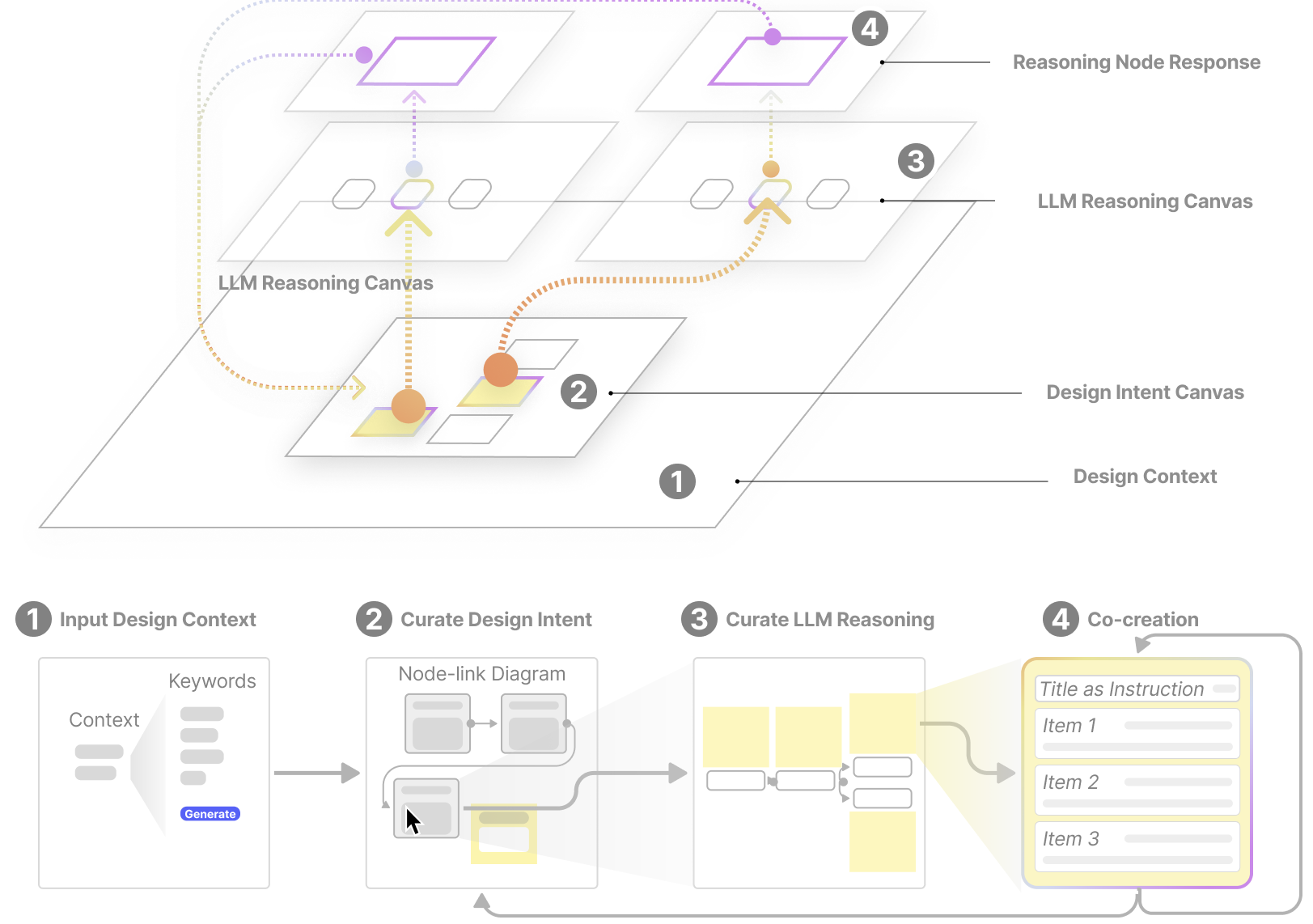}
    \caption{Overview of the curated reasoning workflow in DesignerlyLoop, an LLM-based diagramming system for design intent formulation. The system adopts a two-layer diagrammatic architecture—comprising Design Intent and LLM Reasoning canvases—alongside a set of interaction techniques that enable users to externalize, reorganize, correct, and selectively regenerate reasoning processes.}
    \label{fig:hierarchy}
\end{teaserfigure}


\received{20 February 2007}
\received[revised]{12 March 2009}
\received[accepted]{5 June 2009}

\maketitle




\section{Introduction}
    \input{Sections/1_Introduction}

\section{Related Works}
    \input{Sections/2_RelatedWorks}

\section{Formative Study}
    \input{Sections/3_FormativeStudy_rev2}

\section{System Design}
    \input{Sections/4_SystemDesign}

\section{User Study}

\input{Sections/5_UserStudy}
\section{Findings}
    \input{Sections/6_Findings_rev2}
\section{Discussion}
    \input{Sections/7_Discussion}
    
\section{Conclusion}

\input{Sections/9_Conclusion}

\section{Acknowledgment about the Use of LLM}
The authors would like to acknowledge the use of the generative AI tool in this work. Specifically, \textit{GPT-4o} by OpenAI was utilized to: (1) assist in language refinement, including grammar and style corrections of existing manuscript text, (2) generate R code for data analysis based on our proposed analytical procedures, and (3) generate LaTeX tables from the analyzed data results. Moreover, \textit{GPT-4o} model and \textit{text-embedding-ada-002} model API service was used through Microsoft Azure interface during system implementation. All interpretations, conclusions, and final content remain the responsibility of the authors. 











\begin{acks}
\end{acks}

\bibliographystyle{ACM-Reference-Format}
\bibliography{DesignerLoop,references}


\appendix
    \input{Sections/9_Appendix}

\end{document}

%% file: Sections/0_Abstract.tex
Recent large language models (LLMs) show promise in design tasks, yet a fundamental misalignment persists: design thinking requires iterative intent formulation, while LLMs treat inputs as complete specifications. This challenges design intent formulation, where designers must progressively refine understanding through exploration. 
Existing tools either sacrifice exploratory flexibility for structural stability or leave reasoning implicit, failing to support human-LLM alignment. 
Through a formative study with eight designers, we introduce curated reasoning—enabling designers to explicitly inspect, reorganize, and selectively regenerate LLM reasoning structures. We present \DL{}, implementing this through a two-layer structure separating design intent from LLM reasoning. A study with 20 designers demonstrates that curated reasoning significantly improves design quality and creativity. Our work contributes a novel interaction paradigm for human-LLM alignment, transforming LLMs from content generators into structured reasoning partners in creative design. 


%% file: Sections/1_Introduction.tex
Recent works support the capabilities of large language models (LLMs), such as GPT-5, in addressing diverse design needs~\cite{10.1145/3613904.3642266,asadi_2023_LLMs}: design ideation~\cite{he_2024_ai,chen_coexploreds_2025,shen_ideationweb_2025}, concept development~\cite{tholander_design_2023,anjum_ink_2024,urban_davis_designing_2021,jonsson_cracking_2022,zhou_understanding_2024}, and problem-solving~\cite{10.1145/3613904.3642466, feng_designing_2023, sun_llms_2024}. 
However, a fundamental tension persists in human-LLM collaboration for design: while \emph{design intent formulation} operates through iterative cycles, where understanding emerges from questioning, partial insights, and progressive reframing~\cite{dorst2001problem,schon1983reflective,law2020design}—current \emph{LLM interaction} treat each user input as a discrete, complete task~\cite{holtzman_surface_2021,zamfirescu-pereira_why_2023}. This creates a critical \emph{human-LLM misalignment}: \textit{the paradigm of LLM interaction fundamentally conflicts with the nonlinear nature of design thinking}. 

\textcolor{black}{This misalignment manifests in two aspects. First, LLMs typically produce outputs without exposing the intermediate reasoning process that connect design intent to generated results, making it difficult for designers to identify \textit{where} and \textit{why} the output diverges from their evolving understanding~\cite{subramonyam_bridging_2024}. Second, when designers attempt to correct or refine outputs through follow-up prompts, they must either accept the LLM's hidden reasoning process or restart the entire generation—neither option supports the iterative refinement of reasoning itself. Consequently, designers risk skipping critical thinking process such as planning, execution, and reflection, leading to over-reliance on LLM outputs, weakened independent judgment~\cite{subramonyam_bridging_2024}, and diminished critical thinking~\cite{zhou_understanding_2024,xu_jamplate_2024,angert_spellburst_2023}.}

Structured yet flexible interaction approaches with LLMs has emerged to address such human–LLM misalignment~\cite{subramonyam_bridging_2024,seidel2018autonomous}. One line of work curates AI outputs through chain-based interaction~\cite{wu_ai_2022,arawjo_chainforge_2024}, while another employs diagrammatic or hierarchical structures to scaffold sensemaking~\cite{chen_coexploreds_2025,shen_ideationweb_2025,angert_spellburst_2023,riche_ai-instruments_2025,lin_jigsaw_2024}. However, such structures often stabilize meaning prematurely and implicitly assume reasoning should converge before exploration continues. Conversely, iterative approaches using templates~\cite{xu_jamplate_2024,xu_productive_2025} or prompt artifacts~\cite{lin_inkspire_2025,shi_brickify_2025} supports nonlinear creative process, but leave reasoning implicit with intuitive generation rather than enabling users to explicitly inspect, compare, or revise reasoning chain as context evolves. 
A fundamental tension thus emerges: existing tools either sacrifice exploratory openness for structural stability, or encourage iteration without supporting \emph{explicit curated reasoning}—the ability to inspect, compare, and selectively retain reasoning elements throughout the design process. 
Bridging this gap requires: (1) \emph{iterative evaluation} to continuously assess both human and LLM reasoning as design intent evolves~\cite{subramonyam_bridging_2024}, and (2) \emph{customizable structure} to organize reasoning through different logics. 
Addressing this gap, we ask: \emph{How can curated reasoning support design intent formulation to achieve human-LLM alignment in creative design tasks?}

\hl{
Inspired works on interaction with LLM chains (e.g., \mbox{\cite{subramonyam_bridging_2024,wu_ai_2022,wu_promptchainer_2022,jiang_promptmaker_2022}}) and information exploration (e.g., Sensecape~\mbox{\cite{suh_sensecape_2023}}), 
our system introduces a nested two-layer diagram structure—design intent and LLM reasoning—and a suite of interaction techniques that allow users to externalize, reorganize, correct, and selectively regenerate reasoning (Figure~\mbox{\ref{fig:hierarchy}} and Figure~\mbox{\ref{fig:interface}}). 
}
In a within-subject study with 20 designers, \DL{} supported both the formation of design intent and critical reflection. Qualitative and quantitative results indicated improvements in design intent formulation, creativity, and design quality, as participants systematically engaged with reasoning and actively curated LLM reasoning rather than passively accepting generations. These findings suggest \DL{} that embedding LLMs as curated reasoning can foster more reflective, iterative, and higher-quality human–AI collaboration. 

Our contributions are threefold: \textbf{(1) Identifying alignment challenges:} Through a formative study with eight designers, identifying three core challenges on curated reasoning for human-LLM alignment (Section~\ref{sec:FS-challenges}) and deriving three design guidelines (Section~\ref{sec:FS-DGs}) that inform the development of our two-layer structure; 
\textbf{(2) System implementation}: Developing \DL{}, implementing curated reasoning through a dual-layer structure that separates design intent from LLM reasoning, with interaction techniques for iterative evaluation and customizable organization; 
\textbf{(3) Empirical validation}: Demonstrating the effectiveness of \DL{}'s curated reasoning improves human-LLM alignment, supporting more effective design intent formulation with enhanced creativity and design quality. 

%% file: Sections/2_RelatedWorks.tex
\begin{table*}[]
\begin{tabular}{lp{11cm}}
\multicolumn{1}{l}{\textbf{Term}} & \multicolumn{1}{l}{\textbf{Definition}}  \\ \hline
\textbf{Design Intent}            & \textbf{Evolving understanding of design goals, manifested through multi-level abstractions} (i.e., high-level goals, contextual rationale, concrete constraints) 
\\
\textbf{Reasoning Structure}      & Organized representation of inferential steps connecting intent to output, can be human-generated or AI-generated   
\\
\textbf{Externalization}          & Process of converting internal cognitive processes into persistent, manipulable artifacts 
\\
\textbf{Curated Reasoning} & Explicit inspection, comparison, and selective retention/revision of reasoning elements as design intent evolves 
\end{tabular}
\caption{Term Definitions.}
\label{tab:term}
\end{table*}
To outline our study, we first discuss the definition and practical needs of \emph{design intent}, followed by a review of existing AI-supported graph-based creativity tools. All key terms are defined and summarized in Table~\ref{tab:term}. 

\subsection{Design Intent Formulation}
\subsubsection{Defining Design Intent}
Design intent rarely emerges fully formed. Instead, it unfolds iteratively through sense-making, prototyping, and reflection as designers engage with evolving constraints and materials~\cite{hayes1996framework}. In such exploratory process, intent may manifest as ambiguous metaphors or affective narratives in the early phrase, later refined into clearer propositions~\cite{nancy1993mentalmodel}. HCI tools—such as sketch canvases, versioned annotations, and timeline overlays—can scaffold this gradual externalization and stabilization of intent over time~\cite{ma_sketchingrelatedwork_2023,ding_structuring_2025,suh_luminate_2024,shaer_ai-augmented_2024,shen_ideationweb_2025,chen_coexploreds_2025}.

\subsubsection{Forming Intent from Diverse Supports}
Creative design processes oriented around the articulation and refinement of intent require nuanced system support that accommodates the evolving and situated nature of intent formation. Drawing from cognitive models of creativity and interaction design literature~\cite{riche_ai-instruments_2025,norman1986cognitive,wang2024exploring,ford_reflection_2024,seidel2018autonomous,degen_exploring_2023,IxDF2015Gulf}, we identify three critical dimensions along which design tools must support intent-driven practices:  

    \textit{Non-Linear design loops.} 
    Creative design intent does not progress linearly but evolves through recursive loops involving exploration, reinterpretation, and reformulation~\cite{candy_creative_2020,druga_scratch_2023,ford_role_nodate,zhou_understanding_2024}. Designers fluidly transition between divergent/convergent thinking across research, ideation, and critique phases~\cite{brown2008design,lin_jigsaw_2024}. 
    This loop feature necessitates tools that accommodate:
        (1) Intent reformulation triggered by new constraints, user feedback, or serendipitous discoveries; 
        (2) Stage-agnostic input mechanisms that allow ideation at any point in the process~\cite{ge_scaffolding_2005,hu_designing_2024};
        (3) Concept tracing features and timeline visualization to revisit and branch design paths~\cite{subramonyam_bridging_2024}. 
    Systems should thus enable temporal fluidity, branching, and re-entry, rather than enforce prescriptive workflows. 

    \textit{Diverse representational forms.} 
    Design intent manifest heterogeneously across practitioners even when addressing identical goals. Diverse solutions emerge through varied emphasis on visual semantics, functional aims, user considerations, or prototype fidelity, reflecting distinct value judgments and objectives~\cite{carroll1995scenario,krippendorff2006semantic}. 
    This diversity necessitates interfaces that fluidly adapt to individual designer's evolving representational needs. 

    \textit{Dialogic co-creation.} 
    Building on Schön’s notion of  ``conversation with materials'', design is an iterative dialogue where emerging artifacts influence cognition and action. Intent formation is not solely internal but shaped through back-and-forth interaction with evolving prototypes, sketches, and other intermediaries~\cite{riche_ai-instruments_2025}. 
    Design tools should therefore treat artifacts not just as outputs but as conversational partners—surfacing meaningful responses that invite reflection, reinterpretation, and thematic development. 

\subsubsection{LLM Capabilities for Building Design Around Intent}
The breakthrough of prompting LLMs (e.g. GTP-4o~\cite{openai2024gpt4o}, Deepseek, Claude) exhibit three core capabilities enabling novel approaches to intent-driven design: 
First, LLMs' advanced natural language comprehension allows designers to verbally externalize and refine ambiguous intent through iterative dialogue~\cite{openai2024gpt4o}. This capability transforms linguistic expressions directly into executable specifications, facilitating on-demand tool creation without technical retraining. The recursive interaction pattern supports continuous intent disambiguation via semantic negotiation~\cite{caramiaux2022explorers}. 


Second, as reflective collaborators, LLMs demonstrate cognitive partnership across creative domains~\cite{yuan_wordcraft_2022,cai_pandalens_2024,qin_toward_2025}. 
They enable \textit{bidirectional} sense-making where both human and AI agents progressively adapt their reasoning, surpassing tools requiring unilateral user adaptation~\cite{shi_understanding_2023,ford_reflection_2024}. 

Third, LLMs perform abductive inference—generating plausible hypotheses under uncertainty—through frameworks like ReAct that integrate reasoning with action~\cite{yao2023reactsynergizingreasoningacting}. This capability facilitates contextual intent inference with under-specified goals, consequence anticipation prior to implementation, and comparative solution evaluation. Crucially, LLMs enable analogical adaptation of known forms to novel problems through cross-domain knowledge transfer. This aligns with Peirce's conception of abduction as ``inference to the best explanation''~\cite{peices2025abduction}. By translating abstract concepts into executable specifications while maintaining semantic coherence across iterative refinements, LLMs scaffold design intent as a dynamic cognitive process co-negotiated between human and artificial agents. 


\subsection{AI Graph-Based Creativity and Human-AI Co-creation Tools} 
The fundamental cognitive gap between LLM and designer makes it difficult for designers to align their evolving design process with the structured outputs provided by LLMs. 
LLM often response with direct, objective-driven, and turn-based outputs (e.g., via instructive prompts like “You are a [role]”)~\cite{zhou_understanding_2024,subramonyam_bridging_2024,angert_spellburst_2023,xu_jamplate_2024}. In contrast, design is inherently nonlinear, iterative, and ambiguous, often characterized as dealing with ``ill-defined problems'' due to incomplete or evolving problem information~\cite{simon1971human}. 

Graph-based interfaces resolve this gap by enhancing editability and information representation, thereby supporting nonlinear design workflows. Such interfaces enable refinement, regeneration, and navigation across design iterations \cite{riche2025aiinstruments,shen_ideationweb_2025,chen_coexploreds_2025,xu_jamplate_2024,jiang_graphologue_2023,wu_promptchainer_2022,choi_creativeconnect_2024,lin_jigsaw_2024,lin_inkspire_2025}. These tools extend established visual methods (e.g., mind maps, concept maps, node-link diagrams) that scaffold ideation and systems thinking in design research \cite{cai_designaid_2023,odonovan2015designspace}. By facilitating multi-turn iteration and pipeline composition, they introduce novel LLM interaction paradigms. 
\citet{shokrizadeh_2025_dancing} enables UI ideation through version navigation on a canvas. 
\citet{lin_jigsaw_2024}'s \textit{Jigsaw} system supports cross-modal AI pipeline construction (e.g., integrating ChatGPT and Midjourney). 
\citet{he_ai_2024} employs an LLM-enhanced canvas for group ideation with structured sticky notes. 
Graph structures further serve as intermediary representations for prompt steering by decomposing text prompts into hierarchical elements \cite{yan_xcreation_2023,riche_ai-instruments_2025}. Systems like \textit{AI-Instruments} \cite{riche_ai-instruments_2025} utilize fragmented prompt cards to enable continuous intent refinement. 
However, some inherent limitations of current graph-based approaches constrain their design support capability. Limited controllability and interpretability can erode human trust in AI systems~\cite{liao2023designerly,rai2020explainable,terry2023alignment}.
Moreover, users may become overwhelmed by the accumulation of prompts and AI-generated outputs during extended exploration~\cite{gero2024supporting}.

Human-LLM co-creation shows particular promise during conceptualization. AI agents function as co-creators that introduce varied inputs in parallel, enhancing conceptual diversity~\cite{muller2025genai,houde2025controlling,muller2022framework}. Empirical evidence indicates that even brief exposure to AI-generated ideas stimulates novel conceptual directions~\cite{lavrivc2023brainstorming}. However, persistent concerns regarding authorship ambiguity, over-reliance, and diminished human agency require resolution~\cite{10.1145/3591196.3593364}. 
Scholars have increasingly employed visualized approaches to represent co-creation process, not only increasing model understanding but also preserving user ownership and intentionality~\cite{shen_ideationweb_2025,riche_ai-instruments_2025,angert_spellburst_2023,suh_luminate_2024,xu_jamplate_2024}. 
\textit{IdeationWeb}~\cite{shen_ideationweb_2025}, for instance, empirically applied four distinct human-AI collaboration strategies in design ideas, yet its single-turn generation mechanism still limited user controllability.  
One type of these studies focusing on reasoning processes in such co-creation processes have shown potential to address this issue. \textit{Jamplate}~\cite{xu_jamplate_2024} employs design templates (i.e., Five Whys, Competitive Analysis) to facilitate progressive, stepwise solution refinement and cognitive elicitation. However, its use of fixed node-link structures inadequately captures the evolving nature of design processes, thereby constraining emergent exploration. \textit{CoExploreDS}~\cite{chen_coexploreds_2025} fills this gap by employing node-linked diagram enables track and visualize idea development, capturing logical relationships and conceptual similarities. Furthermore, this work leverage analogical, inductive, abductive and analogical reasoning to guide LLM thinking during ideation~\cite{chen_coexploreds_2025}. 

Nevertheless, these systems still provide insufficient user control over the underlying reasoning processes. Users are left with multiple rounds of what is essentially one-time generated content to continually refine. 
Externalizing reasoning structure enables designers to evaluate LLM cognition and better align outputs with design goals. 
Our work aims to explore a more explicit externalization of reasoning strategies to extend this boundary by enabling bidirectional integration between design representations and LLM reasoning—transforming visual design diagrams into editable interfaces for AI interaction.




%% file: Sections/3_FormativeStudy_rev2.tex

We conducted formative studies focusing on revealing challenges during design intent formulation using LLMs (Section~\ref{sec:fs-study1}) and design guidelines (Section~\ref{sec:FS-DGs}). 

\subsection{Study: Understanding Human–LLM Misalignment}~\label{sec:fs-study1}
This study investigates how designers form design intent using LLMs during early-stage concept development, \hl{with a focus on identifying structural breakdowns between designers' reasoning needs and current LLM interaction paradigms.} \hl{This study is across two stages: (\textit{I}) semi-structured interview with eight design experts (Table~\mbox{\ref{tab:fs-participants}}); (\textit{II}) one-on-one design-activity with three of them (P1, P7, P8 in Table~\mbox{\ref{tab:fs-participants}}) for interaction analysis. All studies are approved by the university IRB.} 

\begin{table}[]
    \centering
    \begin{tabular}{cccc}
       ID  & Design Background & Design Exp & LLM Exp \\
       \hline
        1 & architecture\&product & 8-year & daily; 3-year\\
        2 & architecture\&urban & 8-year & weekly; 3-year\\
        3 & visual communication & 3-year & weekly; 2-year\\
        4 & UIUX\&product & 6-year & daily; 3-year\\
        5 & UIUX\&product\&arch & 8-year & daily; 3-year\\
        6 & visual\&UI & 7-year & weekly; 2-year\\
        7 & product & 5-year & daily; 3-year\\
        8 & product\&service & 8-year & daily; 3-year\\
    \end{tabular}
    \caption{Summary of participants' demographics in the formative studies.}
    \label{tab:fs-participants}
\end{table}
\subsubsection{Participants}
Eight designers (5 female, 3 male; aged 23–30; Table~\ref{tab:fs-participants}) were recruited via purposeful sampling in professional design networks and university design departments. Inclusion criteria required (1) a minimum of two years of design experience and (2) at least one year of weekly LLM use. 
Participants represented a range of design domains, including visual communication, UI\&UX, product, service, architecture, and urban design. 
\hl{From this cohort, three participants (P1, P7, and P8) voluntarily enrolled in the \emph{Stage II design activity}. Participation was uncompensated, and all provided informed consent through an online registration survey prior to the study.} 

From this cohort, three participants (P1, P7, and P8) voluntarily enrolled in the \emph{Stage II design activity}. Participation was uncompensated, and all participants provided informed consent through an online re-registration survey prior to the study.

\subsubsection{Procedure and data.} Our study consisted of two stages: 
\paragraph{Stage I-interview.} 
    \hl{We first conducted semi-structured interview with all eight experts, centering on (1) processes of and approaches on design intent formulation; (2) interaction with LLMs, such as queried questions and provided information; (3) future LLM tool envision. 
Each interview lasted around 30 minutes, recorded and transcribed. 
All participants noted that they get used to inspired by GPT and organizing unstructured intent using graphic tools like FigJam, Miro. These recurring topics formed our focuses in the \emph{Stage II}.} 

\paragraph{Stage II-design activity.} \hl{Before start, a researcher introduced task goal, procedure, and specified tools (graphic tools in \textit{FigJam}~\footnote{a popular graphic tool to organize information with a digital canvas} and GPT-4). The study procedure is as follows: 
    (1) \emph{A design task (30min)} using GPT asked participants to create a design concept and present evolving design intent in on \textit{FigJam}. During design process, we encouraged them think-aloud~\mbox{\cite{ericsson1980verbal}}. 
    After the task, we collected their GPT dialogue records and exported pdf of FigJam board. 
    (2) \textit{A semi-structured interview (15min)} followed, focusing on how GPT helped to shaped design intent formulation and challenges.} 
All on-screen actions, prompt sequences, generated artifacts, verbal protocols, interviews were recorded, audios were transcribed. 
These materials captured formulation, articulation, and revision of design intent during LLM-assisted design activities. 

\subsubsection{Data Analysis.} We conducted an open coding analysis~\mbox{\cite{corbin2014basics}} to analyze the qualitative data. Two researchers independently performed open coding on interview transcripts and design artifacts collected during the design activity, generating data-grounded codes. Coding results were compared and discussed in regular calibration meetings to resolve discrepancies and consolidate a shared codebook. 
A senior researcher reviewed the coding process to support methodological rigor. The finalized codebook was applied to the full dataset. Analytic memos documented coding decisions and translation clarifications. Higher-level categories were developed by aggregating related codes across interviews and behavioral observations. 

\subsubsection{Findings: Three Challenges in Human-LLM Alignment during Design Process}~\label{sec:FS-challenges} 
Participants exhibited three recurring challenges rooted in fundamental limitations of conversational LLM interfaces: 

\paragraph{C1: Reasoning Opacity—Inability to Inspect LLM's Inference Process.}
A foundational challenge in explainable AI is that neural models generate outputs without exposing intermediate reasoning steps~\cite{rai2020explainable,saeed_explainable_2023}. In design contexts, this opacity manifests critically: when LLMs propose directions (e.g., ``use modular components for elderly furniture''), designers cannot discern \textit{whether} the model applied analogical reasoning from precedents, deductive logic from stated constraints, or inductive generalization from user data.

Participants reported that GPT outputs felt like ``conclusions without derivations'' (P5). P1 noted, ``I can't tell if it's making connections I missed or just pattern-matching my keywords.'' When asked to explain reasoning, GPT provided post-hoc rationalizations rather than transparent process traces. This aligns with broader HCI findings that users need access to \textit{reasoning provenance}—not just final recommendations—to evaluate AI suggestions critically~\cite{brachman_building_2025}.

Behaviorally, participants spent 40–60\% of design session time reverse-engineering outputs through follow-up prompts (``why this material?'' ``what assumptions underlie this?''). Even when GPT elaborated, explanations remained \textit{monolithic narratives} rather than \textit{inspectable step sequences}. Designers could not identify which specific inference step caused divergence from their intent, forcing blanket accept/reject decisions.

\textit{Challenge implication:} Systems must decompose LLM reasoning into discrete, inspectable steps that reveal inference methodology—enabling designers to validate process, not just outcomes.

\paragraph{C2: Reasoning-Intent Conflation in Flat Conversational Interfaces.}
Design cognition research distinguishes between \textit{problem framing} (what to design) and \textit{problem solving} (how to achieve it)~\cite{dorst2001problem}. However, current LLM interfaces collapse these layers: design intents (``design sustainable urban furniture'') and supporting reasoning (``elderly users need stability, therefore low center of gravity...'') intermingle in linear chat histories. This conflation creates two problems:
~
First, \textbf{conceptual fragmentation}: P7 maintained separate conversations for high-level goals versus detailed reasoning, noting ``I can't keep design philosophy and technical analysis in the same thread—it becomes unmanageable.'' When design intent evolved (e.g., shifting from ``sustainability'' to ``community engagement''), designers could not isolate and preserve relevant reasoning from obsolete context.
~
Second, \textbf{non-modular revision}: P4 explained, ``If I want to change one assumption midway, I either accept the whole reasoning chain or start over.'' There was no mechanism to modify a specific reasoning step (e.g., revise a constraint) while retaining validated downstream inferences. This mirrors limitations identified in mixed-initiative systems research: effective co-reasoning requires \textit{checkpoint-based refinement} where users validate partial progress before propagation~\cite{horvitz1999principle}.

Participants compensated by externalizing structure in FigJam—creating spatial hierarchies to separate intent-level nodes from reasoning-level details. However, these external scaffolds were disconnected from LLM interactions, forcing manual synchronization.

\textit{Challenge implication:} Systems must architecturally separate design-level intents from execution-level reasoning, allowing independent manipulation of each layer.

\paragraph{C3: Context Isolation Across Connected Design Decisions.}
Design processes inherently involve \textit{distributed cognition}~\cite{Hutchins1995}, decisions are interconnected, with earlier choices constraining or informing later explorations. For example, selecting ``modular design'' as a strategy should inform subsequent material selection reasoning. However, LLMs treat each prompt atomically, unable to leverage reasoning structures from prior, related decisions.

P2 noted, ``Every time I explore a new direction, I have to re-explain the same constraints and context from previous conversations.'' P8 spent 12 minutes of a 30-minute session manually reconstructing background: ``Remember the accessibility requirements I mentioned? And the budget constraints? Now consider...'' Even within single conversations, GPT failed to apply reasoning patterns from earlier turns to analogous later queries.

This reflects a fundamental limitation of conversational AI: \textit{reasoning reusability} requires explicit dependency modeling, not implicit context windows~\cite{zhang2025siren}. When participants attempted to build on previous reasoning (``given the sustainability analysis you did, now evaluate costs''), GPT often regenerated analysis from scratch rather than extending existing logic.

Behaviorally, designers created elaborate external systems: maintaining ``design folders'' with screenshots of earlier GPT outputs, tagging related conversations, and manually synthesizing insights across threads. This cognitive overhead interrupted reflective design flow.

\textit{Challenge implication:} Systems must enable reasoning chains to reference and build upon logic from connected design decisions, not regenerate context atomically.

\subsubsection{Summary: Structural Requirements for Curated Reasoning}
The three challenges converge on a need for \textbf{curated reasoning}—the ability to inspect, organize, and selectively retain reasoning elements throughout the design process. Current LLM interfaces fail because they treat reasoning as:
\begin{itemize}
    \item \textbf{Opaque} (C1): Monolithic outputs without inspectable inference steps
    \item \textbf{Flat} (C2): Intent and reasoning conflated in linear conversations
    \item \textbf{Isolated} (C3): Each prompt processed atomically without leveraging connected decisions
\end{itemize}

\subsection{Design Goals}~\label{sec:FS-DGs}

Synthesizing findings from above studies, we derive three design goals that operationalize the technical requirements identified in Studies:
\textbf{DG1: Externalize Reasoning as Navigable Step Sequences.}
To address reasoning opacity (C1), systems must decompose LLM outputs into discrete inference steps, each represented as an inspectable unit. Rather than presenting monolithic narratives, reasoning should be structured as \textit{chains of nodes} where each node corresponds to one logical step (e.g., identifying a constraint, drawing an analogy, synthesizing sub-solutions). This externalization enables two capabilities: (1) \textit{process validation}—users verify not just conclusions but the logic producing them, and (2) \textit{divergence localization}—when reasoning misaligns with intent, users pinpoint the specific step causing issues rather than rejecting entire outputs. 

\textbf{DG2: Architecturally Separate Design Intent from Reasoning Execution.}
To address intent-reasoning conflation (C2), systems must provide distinct \textit{workspaces} for managing high-level design goals versus detailed inference chains. This separation supports three interaction modes: (1) \textit{intent-level navigation}—users organize conceptual relationships between goals without being overwhelmed by reasoning details, (2) \textit{reasoning-level inspection}—users drill into specific intents to audit or modify supporting logic, and (3) \textit{independent revision}—users alter intents without invalidating reasoning, or refine reasoning steps without restructuring intent hierarchy. 

\textbf{DG3: Enable Context Propagation Across Connected Decisions.}
To address reasoning isolation (C3), systems must allow inference chains to reference and build upon logic from related design decisions. When users explore a new intent that connects to prior work (e.g., ``given the accessibility analysis from earlier, now evaluate costs''), the system should \textit{propagate relevant context} rather than regenerating reasoning from scratch. This requires two capabilities: (1) \textit{dependency modeling}—explicit representation of which reasoning chains inform others, and (2) \textit{selective context injection}—automatically incorporating relevant prior inferences when generating new chains. The goal is to transform LLM reasoning from stateless prompt-response into \textit{compositional sensemaking}.

%% file: Sections/4_SystemDesign.tex
\label{sec:system}
\begin{figure*}[h]
  \includegraphics[width=\textwidth]{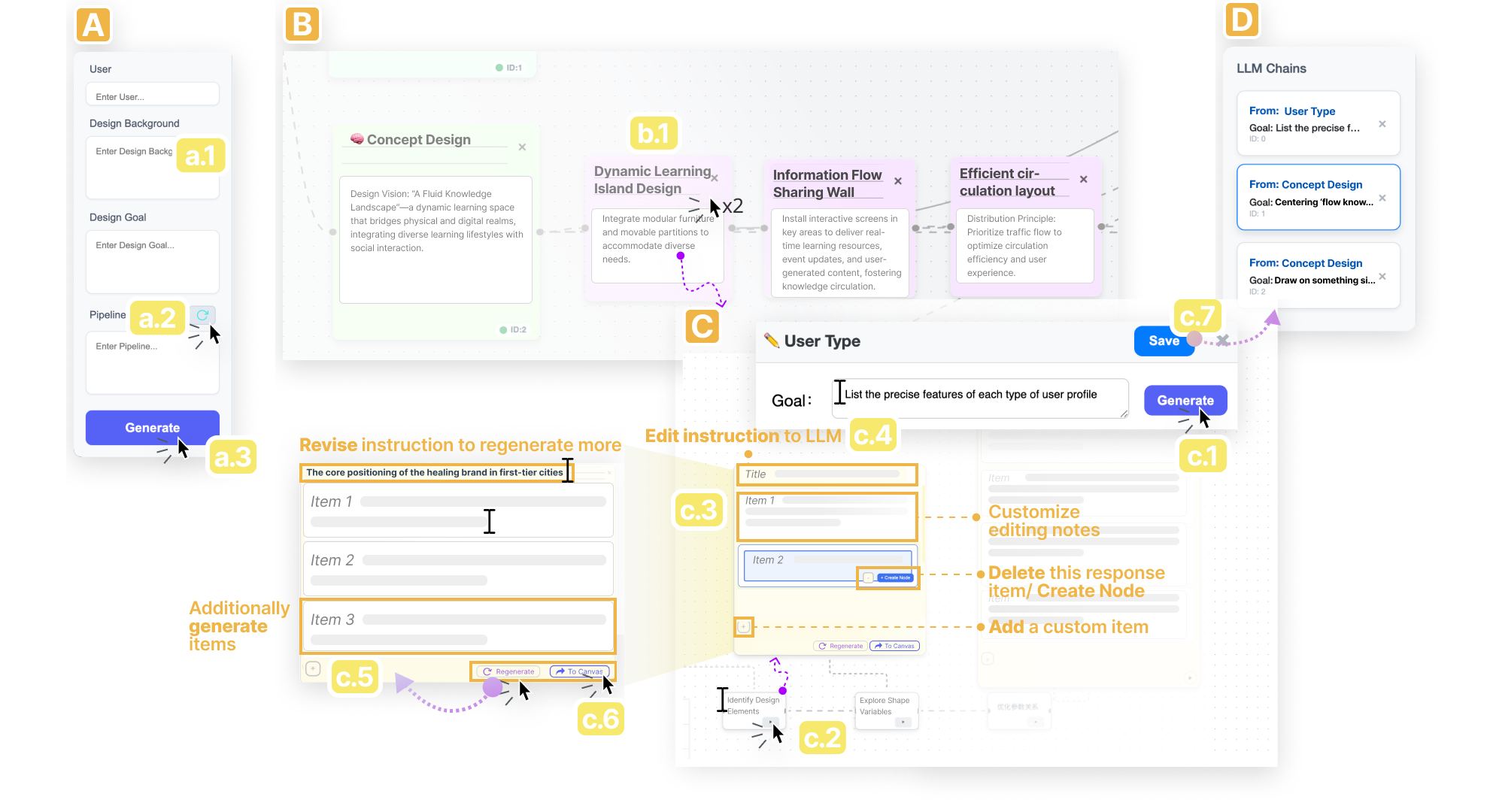}
  \caption
  {
        \DL{} comprises two canvas interfaces: (B) \panelB{}, and (C) \panelC{}.
        Users input design context in \textbf{(a.1)}, generating an editable keyword pipeline \textbf{(a.2)}. Clicking ``Generate'' \textbf{(a.3)} produces a node-link diagram in \textbf{(B)}, supporting customized edits (e.g., add, delete, modify nodes). Double-clicking a node \textbf{(b.1)} opens \textbf{(C)}, where users specify goals \textbf{(c.1)} and obtain LLM-generated reasoning nodes \textbf{(c.2)}. Each node offers multiple design suggestions \textbf{(c.3)} and co-creation functions \textbf{(c.4)}, including content addition, deletion, revision, or regeneration via prompt refinement \textbf{(c.5)}. Finalized outputs can be ``Output to Canvas'' \textbf{(c.6)}, saved \textbf{(c.7)} and checked alongside the canvas \textbf{(D)}. 
    }
  \Description{}
  \label{fig:interface}
\end{figure*}

To support curated reasoning for design intent formulation, we developed \textbf{\DL{}}—a visual node-link diagram system that scaffolds both design intent and LLM reasoning \hl{(as shown in Figure~\mbox{\ref{fig:interface}}). The \mbox{\panelB{}} allows users to organize design intent and cognitive process, collaborated with AI generation. The \mbox{\panelC{}} allows users to engage in reasoning within a LLM chain under specific context on \mbox{\panelB{}}.
This two-layer structure forms a continuous iterative evaluation that enables users to articulate, evaluate and align design intent with LLM. The workflow of this system is shown in Figure~\mbox{\ref{fig:hierarchy}}.} 

\subsection{\DL{} Interaction Design} 
\begin{figure*}[h]
    \centering
    \includegraphics[height=3cm]{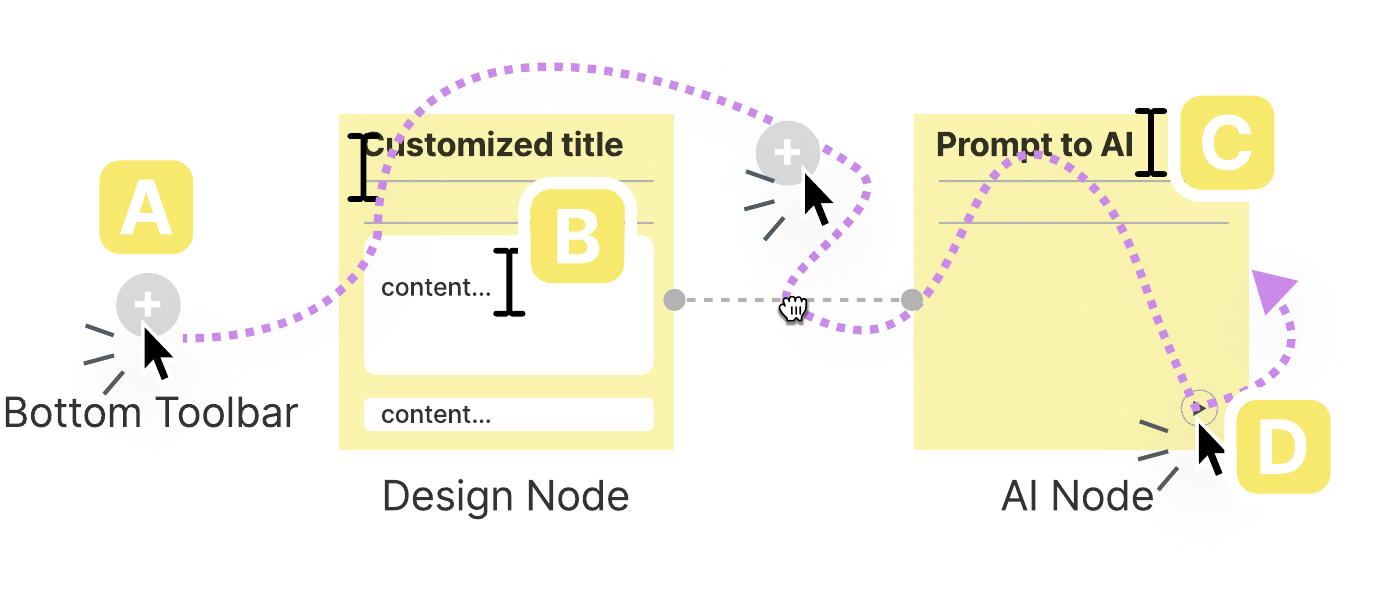}
    \caption{Interaction flow of the Design Intent Canvas integrating \emph{design nodes} and \emph{AI nodes}. Users initiate the workflow by clicking the Add Design Node button to create a design node (A) within the Design Intent Canvas. They then define the node’s title and content blocks (B), which function as structured inputs for the AI node and the Reasoning Canvas. Subsequently, users add an AI node, connect it to the design node, and specify a prompt (C), to produce corresponding outputs (D). 
    }
    \label{fig:aidesignnode}
\end{figure*}
\begin{figure*}[h]
    \centering
    \includegraphics[width=\linewidth]{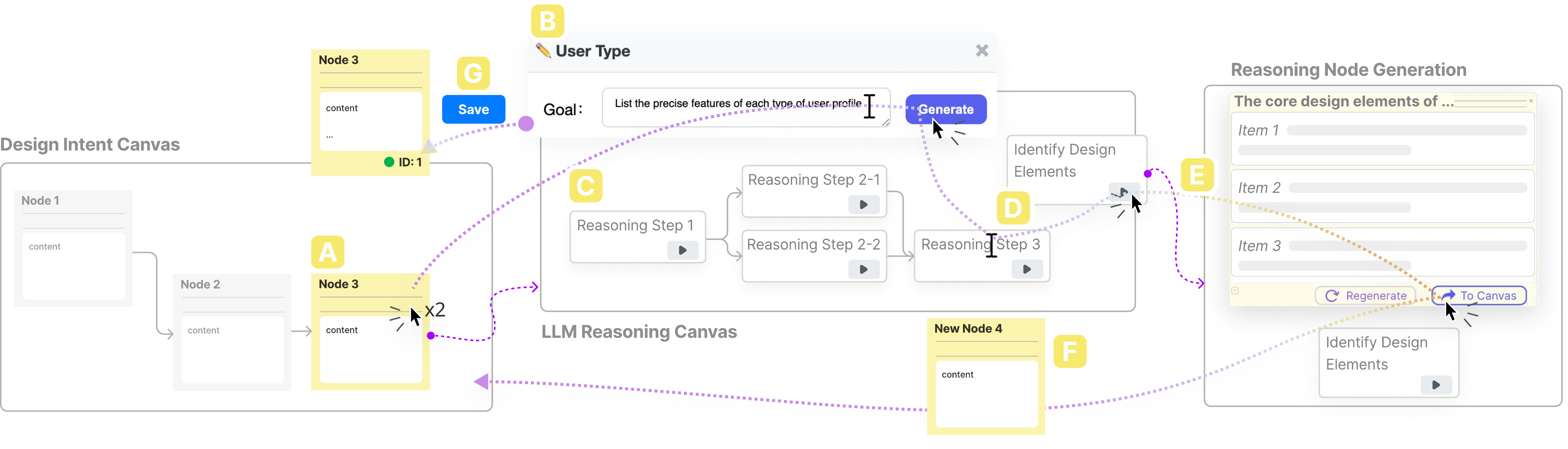}
    \caption{Interaction loop between two-layer structured canvases: user double click a design node (A) on \emph{Design Intent Canvas} to enter \emph{LLM Reasoning Canvas} where user input the query (B) under the current node to generate a LLM reasoning chain (C). This reasoning chain canvas enables user to customize, such as altering node title from \textit{``Reasoning Step 3''} to \textit{``Identify Design Elements''} (D) and generate output in a pop-up Reasoning Node Generation (E) according to this title. Once user click ``To canvas'' would add such output to the main Design Canvas. Saved reasoning chain appears an icon on design node (G). }
    \label{fig:userflow}
\end{figure*}

\subsubsection{\panelB{}}\label{sec:panelB} 
The workflow of this system begins with a entry panel (Figure~\ref{fig:interface}A), that transforms a user's high-level design goal into a concrete, executable design pipeline. 
~
After users specify the design background and goals (Figure~\ref{fig:interface} a.1, a.2), the system 
produces a series of \emph{design nodes}-a structured diagrammatic nodes representing design steps and associated content in \panelB{} (Figure~\ref{fig:interface} b.1). These generated nodes serving as the initial entry for design intent formulation and human--AI co-reasoning.


\paragraph{\hl{Nonlinear Intent Articulation and Refine [\textbf{C2}, \textbf{C3}]}} 
\panelB{} supports design intent formulation by  externalizing \emph{design nodes} within a canvas-based diagramming interface (Figure~\ref{fig:interface}B). Each \emph{design node} supports editable titles and segmented content blocks. User can flexibly employ these nodes to articulate initially unstructured intent, such as problem framing, constraints, strategies, and evaluation criteria. 
~
Users can also add, remove, reorder, or interconnect multiple \emph{design nodes}, as well as invoke AI-generated content through dedicated \emph{AI nodes}, to override, iterate upon, or supplement existing diagram structures. The basic interaction flow is shown in the Figure~\ref{fig:aidesignnode}. 

\hl{Content within \mbox{\panelB{}} collectively functions as a dynamic prompt template, which subsequently guides AI generation within the \mbox{\panelC{}}. This design addresses \textbf{DG2} by externalizing and aligning design intent with LLM’s internal inferences, and \textbf{DG3} by supporting nonlinear and iterative refinement across interdependent design decisions).}

\subsubsection{\panelC{}}
\label{sec:panelC} 
\begin{figure*}
    \centering
    \includegraphics[width=\linewidth]{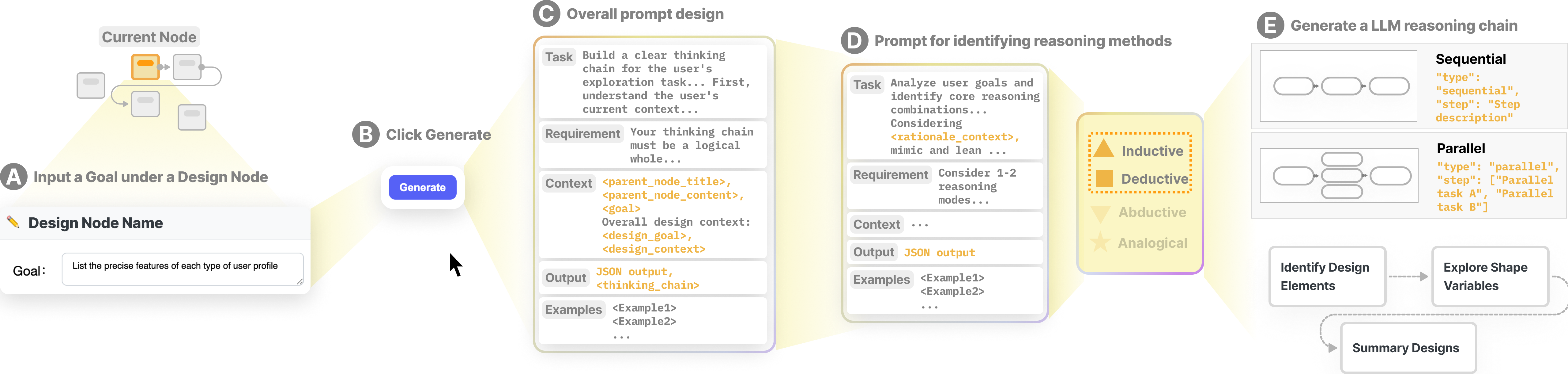}
    \caption{Pipeline for LLM chain generation by the designer on \panelC{}. Double-clicking a design node opens \panelC{}, where users (A) input a specific ``goal'' for in-depth exploration under the current \emph{design node}; (B) click ``Generate'' to (C) map the input into structured prompts, including task, requirements, context, output, and examples; and (D) produce an \emph{LLM chain} that generates both sequential and parallel chain structures with prompt requirements and output formats.}
    \label{fig:pip-goal>chain}
\end{figure*}

\begin{figure*}
    \centering
    \includegraphics[width=0.77\linewidth]{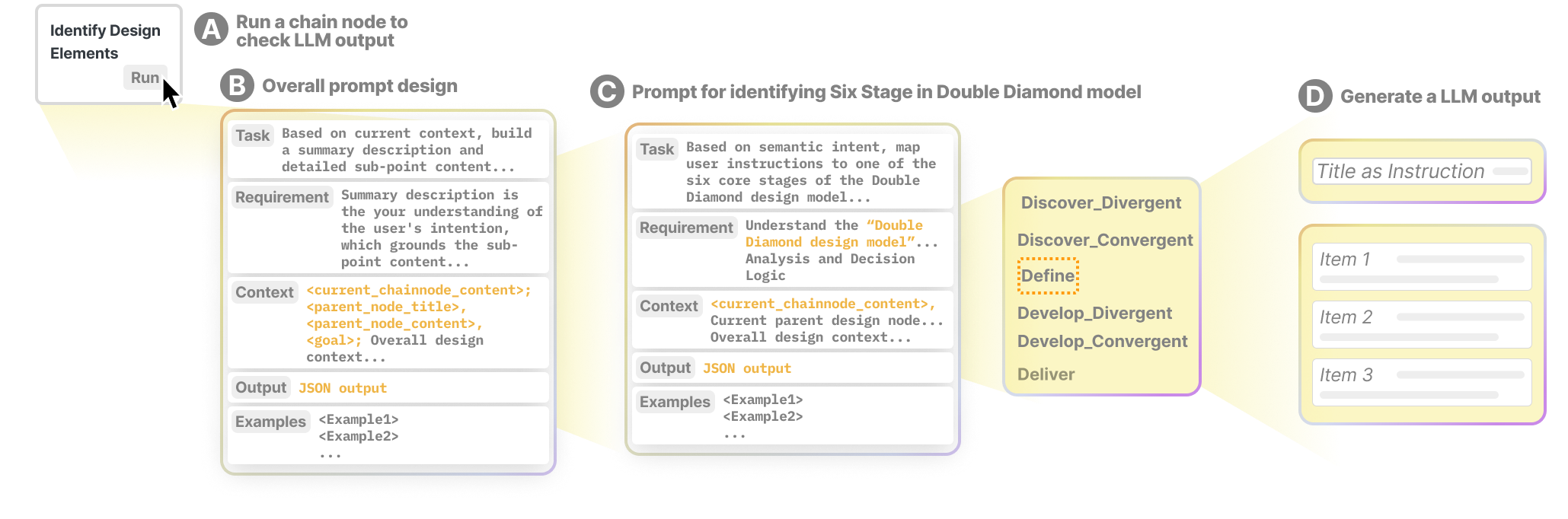}
    \caption{Pipeline for LLM output generation within each chain node on \panelC{}, where users (A) ``Run'' a chain node to obtain the initial LLM output. The system then (B) applies an LLM prompt to identify the corresponding design stage in the Double Diamond model (Appendix~\ref{apx:prompt-sixdesignstages}) and (C) applies another prompt to specify the reasoning method(s) (Appendix~\ref{apx:prompt-fourreasonings}). Finally, (D) the system generates the refined LLM output accordingly.
    }
    \label{fig:pip-chainnode_output}
\end{figure*}
~
By double-clicking a \emph{design node}, users open a secondary-level canvas, \panelC{}, \hl{to explore a specific problem (Figure~\ref{fig:interface}C). Its core function is to generate a structured, multi-step ``LLM thinking chain'' along with rationales that guide the user's creative process. } 
~

\paragraph{\hl{Interactive Hypothesis Steering [\textbf{C1, C2}]}} 
Within this canvas, \hl{users articulate a hypotheses or question, and the system infers reasoning chain (Figure~\mbox{\ref{fig:interface}}C) based on the existing content and structural context in \mbox{\panelB{}}.} 
Users specify a ``goal,'' ``hypothesis,'' or ``question'' in natural language (Figure~\mbox{\ref{fig:userflow}}B), which the system \hl{automatically translates into an editable and reorganizable chain of reasoning nodes generated by the LLM (Figure~\mbox{\ref{fig:userflow}}C).} 

\begin{figure*}[h]
    \centering
    \includegraphics[height=1.2cm]{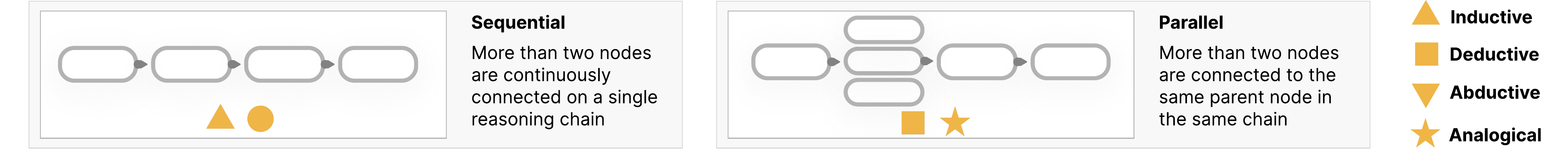}
    \caption{Generated reasoning chains on \panelC{} is sequential or parallel. Each chain considers a primary and secondary reasoning method, selected from inductive, deductive, abductive, or analogical reasoning.}
    \label{fig:twochains}
\end{figure*}

 To generate deeper reasoning, 
 \hl{the system uses an LLM-powered classifier to analyze the user's goal and categorize it into one or two core reasoning methods from predefined four—inductive, deductive, abductive, or analogical (Appendix~\mbox{\ref{apx:prompt-fourreasonings}}), inspired prior study \mbox{\cite{chen_coexploreds_2025}}.} The pipeline is shown in Figure~\ref{fig:pip-goal>chain}. 
 Meanwhile, the system distributes a design stage in \emph{double diamond model}~\mbox{\cite{c0cdf5de-f583-35bb-a33b-a532ccb12896}} for each \emph{LLM chain node} for further human-LLM alignment in \emph{Nodel Level} (Figure~\ref{fig:pip-chainnode_output}). 
 These design stages and reasoning methods are clearly identified in system prompts (Appendix~\ref{apx:prompt-fourreasonings} and~\mbox{\ref{apx:prompt-sixdesignstages}}). Drawing on the dynamic context and prompt templates defined in \mbox{\panelB{}}, this generation produces a multi-step reasoning chain with either sequential or parallel structures (Figure~\ref{fig:twochains}), 
 thereby externalizing LLM’s reasoning process. 



\paragraph{\hl{Reasoning Alignment: Chain Level [\textbf{C1, C2}]}}
\hl{The generated reasoning chain also enables revise and iterate to support curation (Figure~\mbox{\ref{fig:interface}} c.2). 
Each \emph{reasoning step node} supports revisable titles and regenerate according to revised titles. Additionally, user can add, remove, reorder, interconnect or disconnect these nodes. 
~
Clicking ``run'' button on each \emph{reasoning step node}, user enter the following \emph{reasoning alignment} on \emph{node level}. 
} 


~
\paragraph{\hl{Reasoning Alignment: Node Level [\textbf{C1, C3}]}}
Each LLM chain node can generate a multi-dimensional output in a pop-up panel (Figure~\ref{fig:userflow}E) that supports human-LLM alignment (Figure~\ref{fig:interface} c.3). In such alignment, system supports users to add, delete, revise, annotate, and reorder nodes (Figure~\ref{fig:interface} c.4): \emph{Addition}—introducing new ideas or contextual details, which the LLM integrates into supplementary nodes; \emph{Deletion}—removing irrelevant nodes, prompting the LLM to adapt reasoning structure; \emph{Revision}—adjusting wording or scope, leading the LLM to regenerate content; and \emph{Prompt refinement}—reformulating inputs to explore alternative trajectories. 
~
Users can also regenerate content by editing the node title and clicking ``Regenerate'' (Figure~\ref{fig:interface} c.5).

Once refined, each sub-point response under a \emph{reasoning node} can be externalized as sticky notes in \panelC{} via ``Create Node'' (Figure~\ref{fig:interface}~c.4) and each whole response can be exported to \panelB{} via ``Output to Canvas'' (Figure~\ref{fig:interface}~c.6), or saved for later iteration and comparision (Figure~\ref{fig:interface}~c.7). \hl{Through saving feature, user can compare of reasoning differences between versions.} \hl{For instance, user can dive into different perspectives by creating two parallel \emph{design nodes}, A and B, with the same title and content in one diagram in \mbox{\panelB{}}, then create two different reasoning chains within the node A and B.} 

\paragraph{}


Collectively, these mechanisms address \textbf{DG1} by externalizing LLM reasoning processes and making them transparent and controllable to users. At the chain level, interactive steering and alignment ensure that generated reasoning structure remain logically coherent and consistently aligned with design intent, thereby fulfilling \textbf{DG2}. At the node level, fine-grained control over individual reasoning steps further supports \textbf{DG3} by enabling iterative refinement and comparative exploration across alternative reasoning structures. 

\subsubsection{Supporting Reasoning Reuse through Explicit Linking and User-Controlled Context Composition.} 
\begin{figure*}[h]
        \centering
        \includegraphics[width=\linewidth]{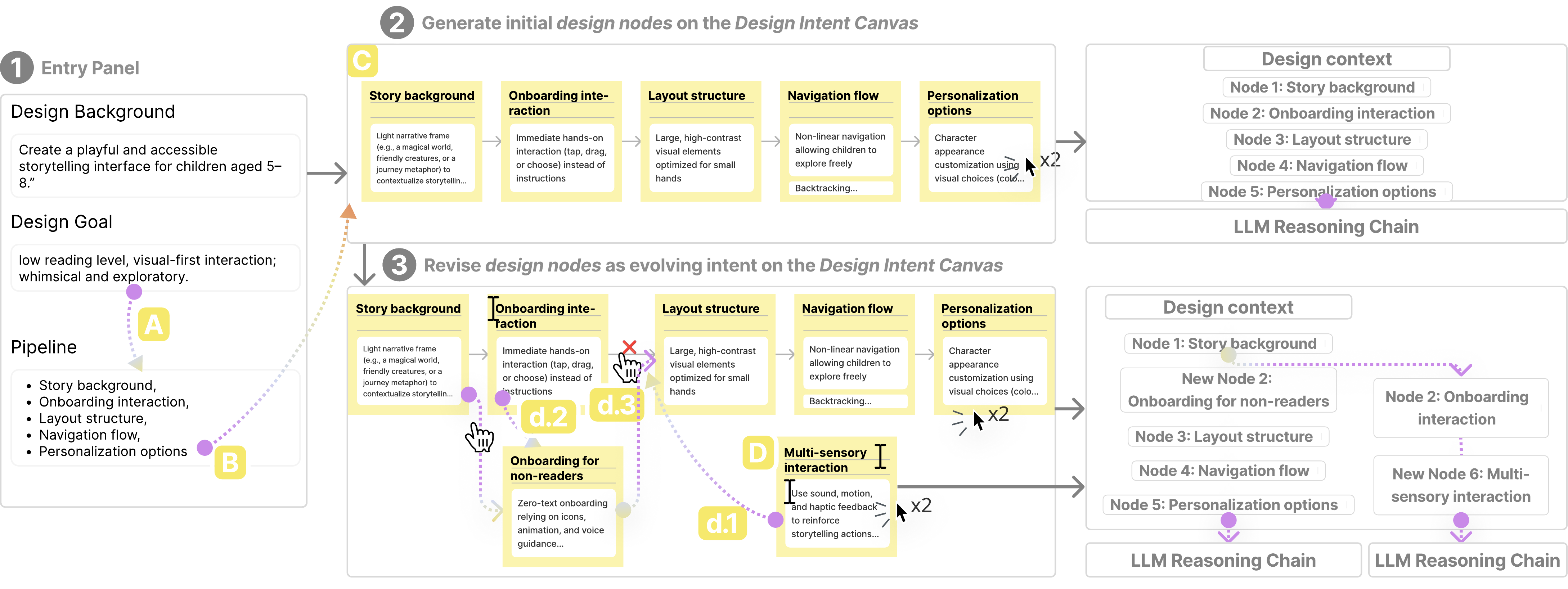}
        \caption{User scenario of Lin's walkthrough, showcasing that revised structure of design node as dynamic input context impacts the subsequent generation of LLM reasoning chain. When users double-click on Node 5 and New Node 6 respectively to enter the LLM reasoning chain, there are two different branches depending on the different upstream contents (Lower right).}
        \label{fig:userscenarios}
    \end{figure*}

To address the challenge of fragmented reasoning (C3), \DL{} does not attempt to automatically infer or propagate relevant context. Instead, it treats prior reasoning as an explicit, reusable artifact that can be selectively reintroduced into subsequent interactions. Users can link, duplicate, or reference design nodes across canvases, making dependencies between intent-level decisions and underlying reasoning visible and inspectable. 
~
Figure~\ref{fig:userscenarios} shows one case in which upstream content on design nodes was restructured after revising for further LLM generation. 

When generating new AI responses, \DL{} composes prompts by incorporating user-selected upstream design nodes (e.g., design titles, summarized rationales, or validated outputs) as contextual constraints rather than implicit background. 
This design emphasizes user-controlled context composition over automated relevance inference, to reinterpret or discard prior reasoning. 
\hl{This explicit linking mechanism directly addresses \textbf{DG3} by maintaining coherence and context propagation across nonlinear, iterative design processes.} 




\textbf{Example User Walkthrough.}
To illustrate this process in \DL{}, we present a walkthrough scenario based on an interface design (Figure~\ref{fig:userscenarios}). Designer Lin explores layout strategies for an interactive storytelling app for children.

The session begins in \emph{Entry Panel} (Figure~\ref{fig:userscenarios}A), where Lin defines a high-level design brief: ``Create a playful and accessible storytelling interface for children aged 5–8.'' They specify preferred goals and constraints in the \textit{``Design Goal''} field (e.g., ``low reading level,'' ``visual-first interaction'') and tag the style as ``whimsical and exploratory.'' The system uses this input to generate a sequenced design pipeline, including: \textit{(1) story background, (2) onboarding interaction, (3) layout structure, (4) navigation flow, (5) personalization options} (Figure~\ref{fig:userscenarios}B). 

Next, the user moves to \panelB{}, where the pipeline is visualized as editable \emph{design nodes} (Figure~\ref{fig:userscenarios}C). After inspecting and refining the structure, Lin feel lacking a specific interaction, thus adding a node for ``multi-sensory interaction'' that was absent in the generated pipeline (Figure~\ref{fig:userscenarios}D), then insert it between two nodes (Figure~\ref{fig:userscenarios}~d.1). 

When deeper reasoning is required, Lin found a refined idea, then duplicate ``Onboarding interaction'' node and revise the new one to ``Onboarding for non-readers'' (Figure~\ref{fig:userscenarios}~d.2) and reconnect the links (Figure~\ref{fig:userscenarios}~d.3) to form two branching links. 
~
Lin double-clicks the relevant node, triggering \panelC{}. The system generates a multi-step LLM reasoning chain, with suggestions such as ``use audio prompts,'' ``add avatar-based guidance,'' and ``progressive disclosure of functionality.'' Lin evaluates, reorders, and revises steps, updating the underlying reasoning graph. 

Throughout the session, Lin iteratively switches between panels, refining goals (entry panel), restructuring the conceptual map (\panelB{}), and interrogating or editing LLM outputs (\panelC{}). This dynamic interplay fosters not only idea generation but also reflection, traceability, and cognitive alignment across the design process.

\subsection{System Implementation}
\begin{figure*}
    \centering
    \includegraphics[width=\linewidth]{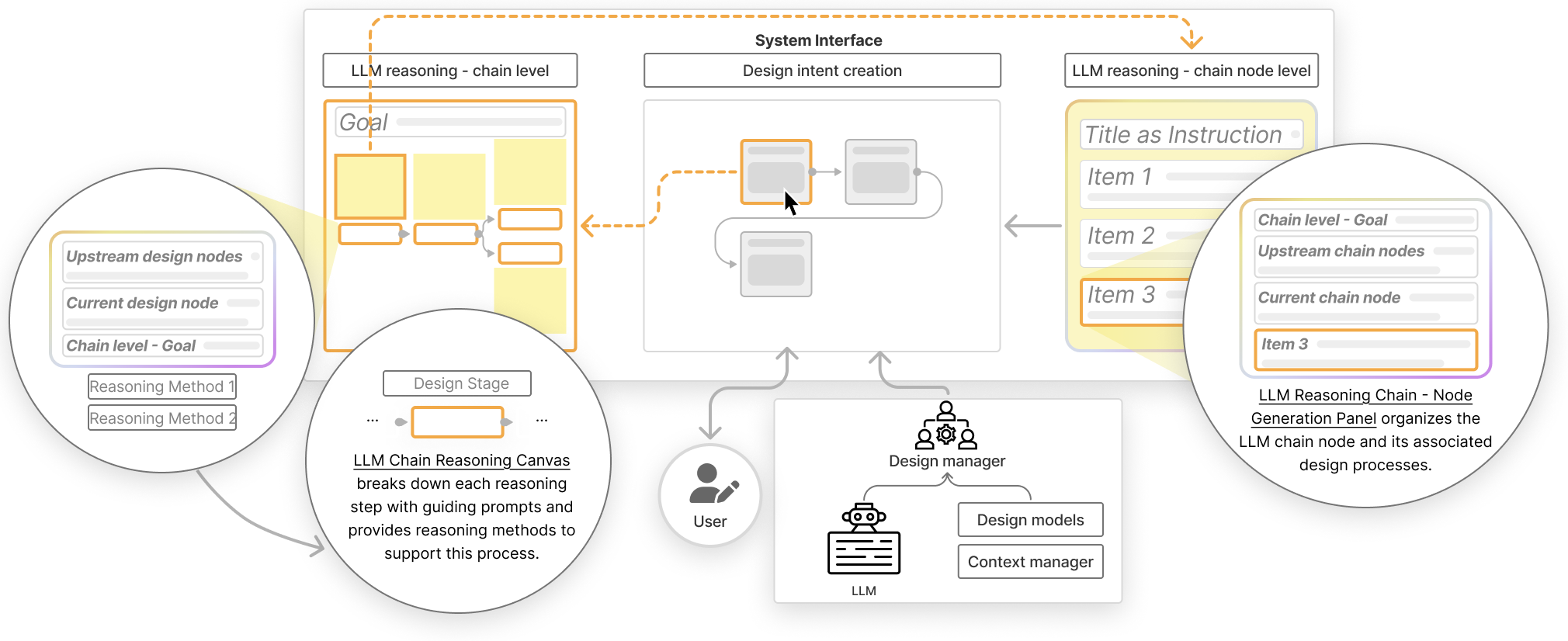}
    \caption{Technical framework of \DL{}, indicating interaction workflow includes three main steps: (1) generating design intent nodes with customized node-linked diagram, curating LLM chain reasoning in (2) chain level and (3) chain node level.}
    \label{fig:tech-arch}
\end{figure*}

\textit{DesignerlyLoop} is implemented as a graph-based system in which both design processes and LLM reasoning chains are represented as structured node–link data (Figure~\ref{fig:tech-arch}). 
A unified graph schema allows the frontend to render non-linear design trajectories while enabling the backend to interpret and manipulate the same structure for multi-step reasoning. To support this architecture, the system includes two core modules: 
a \textit{Context manager}, which maintains the graph-structured representation of all design and reasoning nodes and composes node-local context into LLM prompts (Figure~\ref{fig:pip-goal>chain} and ~\ref{fig:pip-chainnode_output}); and a \textit{Design models} module, which stores reusable process templates (e.g., double-diamond pipelines) that seed and scaffold the design canvas shown in Figure ~\ref{fig:tech-arch}. Together, these components provide a shared computational substrate through which the system can generate, expand, and iteratively refine both human-authored design reasoning and LLM-generated reasoning chains.

~\label{sec:systemimplementation}
The \DL{} system is implemented as a web-based application, where the frontend is built with Vue 3~\footnote{\url{https://vuejs.org/}} in JavaScript, and the backend is developed in Python using the FastAPI framework. Specifically, we utilize the Vue Flow~\footnote{\url{https://vueflow.dev/}} as the very foundation for frontend interface. And for backend, it is implemented using FastAPI~\footnote{\url{https://fastapi.tiangolo.com/}} framework, with Uvicorn as the server. We use LangChain~\footnote{\url{https://www.langchain.com/}} to build the whole pipeline of API-based LLM service, integrating the \textit{gpt-4o} model and the \textit{text-embedding-ada-002 } model through Microsoft Azure OpenAI together with ChromaDB~\footnote{\url{https://docs.trychroma.com/}}.

%% file: Sections/5_UserStudy.tex
To validate the effectiveness of curated reasoning structure in \DL{}, we conducted a within-subjects experiment with 20 professional designers. 



\begin{figure*}
    \centering
    \includegraphics[width=0.99\linewidth]{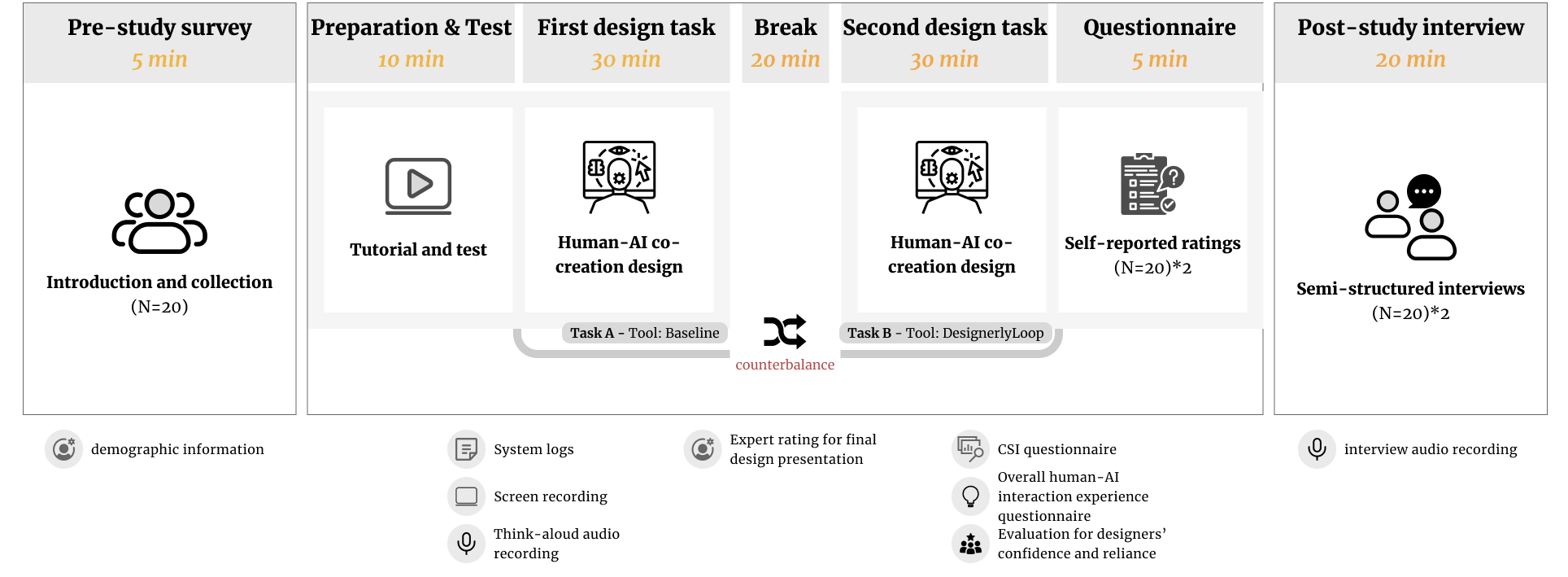}
    \caption{User study procedure for the design task comparing \DL{} and the baseline system.}
    \label{fig:studyprocedure}
\end{figure*}

\subsection{Participants} 
We recruited 20 professional designers with experience in LLM-assisted design and \hl{diversified design expertise} in industrial and art design (Appendix, Table~\ref{tab:participants}). Participants were recruited through purposive sampling via posts on social media platforms (Xiaohongshu, Weibo) and professional design networks (Xiaohongshu, Wechat), targeting individuals with prior design experience using LLMs. Screening collected (1) demographics (age, gender, profession), (2) design experience (duration, field), and (3) GenAI tool experience (types, use cases). \hl{Inclusion criteria are at least (1) one year weekly LLM tool use experience (2) two year professional design experience.} Participants received \$22 USD for a 120-minute study. The study received approval from the university’s IRB.

\subsection{Study Design}
\subsubsection{Goal and Method} We conducted a within-subjects study with 20 designers, each completing two conceptual design tasks using \DL{} and a baseline system. 
 The comparison isolates the contribution of curated reasoning—by holding representational diagramming format and generative AI capability.

\subsubsection{Baseline Rationale}~\label{apx:baseline}
\hl{To isolate the effect of curated reasoning, we designed a baseline system that matched \emph{DesignerlyLoop} in LLM capability, prompt content, and diagramming affordances, differing only in how reasoning structures were externalized and manipulated during interaction. 
Both conditions used the same underlying LLM, identical prompt templates, and the same turn-based generation mechanism. In both systems, AI outputs were generated only upon explicit user invocation, and no additional reasoning depth, hidden chain-of-thought prompting, or intermediate inference steps were introduced in either condition. 
The baseline provided an integrated node-based diagramming canvas with basic AI node generation, allowing participants to incrementally develop and revise design intent within a unified workspace. AI nodes in the baseline were generated based solely on the currently selected node content and the shared prompt, mirroring standard turn-based LLM interaction.
In contrast, \emph{DesignerlyLoop} enabled participants to explicitly curate and reorganize relationships among reasoning nodes, which could then be referenced as structured context during subsequent generations. Importantly, this structure did not introduce new generative functions or increase model capability; rather, it altered how existing content was selected, composed, and interpreted as input context.

We intentionally did not use an external baseline such as GPT plus FigJam. Pilot studies indicated that switching between disjoint conversational and diagramming workspaces introduced substantial interaction overhead and disrupted participants’ reasoning continuity. Our baseline therefore represents a strong, integrated comparison condition, ensuring that observed differences can be attributed to the presence or absence of curated reasoning rather than workspace fragmentation or modality switching.}

\subsubsection{Study Design}
The independent variable was the system used (\DL{} vs. baseline). 
Dependent variables included (1) self-reported SUS (System Usability Scale), (2) AI interaction experience (controllability, collaboration, trust, cognitive load, and enjoyment), (3) agency (artists' self-confidence, AI reliance), (4) creativity support (via the Creativity Support Index), and (5) participants' creative design quality.

\subsection{Procedure}
    \subsubsection{Introduction}  
    Participants signed an informed consent and were introduced to the study context and procedure (as shown in Figure \ref{fig:studyprocedure}). Subsequently, participants watched the tutorial slides with examples for either \DL{} or the baseline system, then lets the participant explore (10 min). 
    
    \subsubsection{Design Tasks} 
        Following the within-subjects design, each participant completed two 30-minute design sessions using both systems. 
        \hl{Participants were asked to select one design problem from a set of six open-ended design problems spanning diverse yet familiar domains, including public space, carbon emissions, caregiver–patient relationships, gamified learning, visual brand recognition, and remote collaboration (see Appendix~\mbox{\ref{apx:designtasks}}). 
        All six design problems were intentionally framed as general, method-agnostic challenges, supporting solution development through a wide range of design approaches across disciplines. 
        Each problem was framed by a single open-ended question (e.g., ``How can design improve people’s experience of waiting in public spaces?'') without domain-specific technical constraints. 
        Allowing participants to select a task aligned with their prior knowledge helped reduce confounds introduced by domain learning and ensured that participants could focus on conceptual reasoning rather than problem comprehension. 
        To preserve comparability across conditions, each participant used the same design topic and conceptual focus for the paired \textit{\mbox{\DL{}}–baseline} conditions.} 
        During each session, participants constructed a node-linked diagram consisting of approximately 5–10 design nodes narrating their design intent. 
        To counterbalance potential task bias, half of participants started with the baseline system and half with \DL{} (Figure~\ref{fig:studyprocedure}). 
        To minimize interference between two design sessions, there is a 20-minute break between the two tasks. Think-aloud protocols were encouraged during the tasks~\cite{ericsson1980verbal}. \hl{Once complete each task, we asked participants to upload produced artifacts, including all canvases and tool interface. 
        }

    \subsubsection{Post-Study Feedback}
    Participants completed questionnaires and semi-structured interviews over 20 minutes. Interviews addressed the following five aspects: 
        (1) usages and experience on design processes using \textit{\DL{}} and \textit{baseline}; 
        (2) differences between these two LLM tools, focusing on aspects such as creative stimulation, collaboration experience, and AI performances; 
        (3) evaluation of \DL{}’s features (curate reasoning; design intent; customized iteration using graph); 
        (4) challenges encountered and impressive moments when using \DL{}; and 
        (5) participants’ willingness in using \DL{} for conceptual design in the future, along with suggestions for improvement. \hl{All sessions and interviews were screen- and audio-recorded for further analysis. During the whole process, we collected data, including produced visual artifacts (content and diagrams in the canvas), text-based LLM generation, interaction for subsequent analysis.}
    
    Last, participants were asked to complete self-report questionnaires in five minutes using a 7-point Likert scale \cite{joshi2015likert}, to evaluate their design processes with AI. These included the Creativity Support Index (CSI) questionnaire \cite{rodrigues_creativity_2023}, a questionnaire assessing overall human-AI interaction experience (i.e., controllable, transparent, cognitive load, collaboration, trust) (Appendix \ref{apx:overallhuman-AI}, Table~\ref{tab:overallhumanAI}), and a scale reporting designers’ confidence and reliance (Appendix \ref{apx:agency}). Participants also rated their creative design outcomes based on two metrics: novelty ($N$) and usefulness ($U$) (Appendix~\ref{apx:selfrated-outcomes}). 
    Additionally, participants were asked to articulate their design concepts and describe outcomes for both tasks in a structured form, in preparation for subsequent expert evaluation using the same assessment criteria as self-rated outcome (Section \ref{sec:expertevaluation}).

\subsection{Data Analysis}

Quantitative comparisons within participants began with Shapiro–Wilk tests for normality. Normally distributed differences were analyzed using paired-sample t-tests, non-normal differences with Wilcoxon signed-rank tests. Pearson correlations were used for normally distributed data and Spearman correlations when assumptions were unmet. Expert ratings were assessed for inter-rater consistency using Kendall’s W test~\cite{KendallsWtest} (Section~\ref{apx:expert-consistencytest}).  

Qualitative interview data were thematically coded using an inductive–deductive approach~\cite{jennifer2006ind-ded} to complement and explain the quantitative findings. First, all audio recordings were transcribed verbatim and verified for accuracy. Identifying information in all data and artifacts was removed, and each transcript was assigned a pseudonym. Transcripts, along with captured interaction logs and screenshots, were imported into NVivo to facilitate systematic coding and maintain an audit trail. 
~
Second, two researchers independently conducted line-by-line open coding~\cite{strauss1998basics} on an initial purposively selected subset of data designed to maximize variation in usage patterns. During this stage, coders documented analytic memos capturing emergent ideas, questions, and provisional interpretations. 
Third, the coders met regularly to compare codes, resolve discrepancies through discussion, and iteratively construct a shared codebook~\cite{macqueen1998codebook}. Each code entry included a concise label, a clear definition, inclusion/exclusion criteria, and exemplar quotations or interaction snippets. Successive drafts of the codebook were versioned and retroactively applied to earlier transcripts when new codes were added. Interrater agreement was assessed on a sample of transcripts using percent agreement and Cohen’s $\kappa$~\cite{miles1994qualitative}, with remaining disagreements resolved through consensus discussion.
~
Finally, the finalized codebook was applied deductively to the remaining dataset. Codes were clustered into higher-level themes and subthemes through collaborative mapping sessions. The resulting main themes and subthemes, reflecting user benefits, human–AI collaboration experiences, and creative outcomes, are reported in Section~\ref{sec:findings}, accompanied by representative quotations.

%% file: Sections/6_Findings_rev2.tex
\label{sec:findings}

To evaluate \DL{} against the baseline, we assessed system effectiveness for system usability for design intent formulation (Section~\ref{sec:finding-benefit}), and creativity supports and outcomes (Section~\ref{sec:finding-creativity}). We also reported a perceived interaction experience and agency in Section~\ref{sec:findings-perception}.

\subsection{How does \DL{} Support Design Intent Formulation?}~\label{sec:finding-benefit}
\subsubsection{Supporting for Design Intent Formulation with High Usability}
\hl{Rather than treating intent as a transient linguistic specification embedded in conversational context, \mbox{\DL{}} reconfigures design intent as an externalized, persistent, and revisable epistemic object. Interview data suggest that this shift directly addresses the misalignment between linear LLM interaction paradigms and designers’ non-linear, evolving cognitive structures.} 

According to the SUS standard, \DL~($M_{DL} = 77.6$, $SD_{DL} = 13.0$) achieved usability score above the acceptance threshold of 70-``Good'' level, the baseline system's score ($M_{baseline} = 69$, $SD_{baseline} = 19.1$) achieved the ``OK'' level. 
The difference was significant, $t(19) = 2.34$, $p = .030$*. 
\hl{We found that three mechanisms through which \mbox{\DL{}} structured, stabilized, and rendered design intent actionable:} 
\textbf{Coherent Process Supports via Explicit and Inspectable Units.}
\hl{Participants described how \mbox{\DL{}} mitigated human-LLM misalignment by externalizing intent into persistent, inspectable nodes that remained available across iterations. 
Rather than relying on memory or re-prompting, participants treated design nodes as stable anchors for their evolving intent. As P3 explained, \textit{``Without this, a new piece of information easily pulls my attention away, and I forget what I was thinking before.''} By making intermediate reasoning visible and spatially persistent, the system allowed intent to remain coherent even as new ideas emerged.}

Participants further emphasized that nested two layers supported this stabilization. 
P18 noted that \textit{``the iterative evaluation between two canvases made it easier to judge the overall logic and coherence of the solution''} (P18) This recursive movement between overview and detail enabled participants to maintain epistemic continuity. 

\textbf{Localized Reasoning Inspection Enables Intent Refinement.}
\hl{\mbox{\DL{}} enabled participants to refine intent by intervening directly in localized reasoning steps. Instead of accepting or rejecting entire AI-generated outputs, participants inspected and modified specific reasoning nodes, aligning AI contributions with their own evolving understanding. } 
~
Participants reported that access to multi-step reasoning made them more critical and intentional in engaging with AI suggestions. As P9 noted, \textit{``I became more willing to adjust its outputs because I could see how it thought through several steps.''} P3 similarly highlighted how direct regeneration and manipulation at the node level helped \textit{``make my thinking more tightly connected.''}

Importantly, participants framed this capability as supporting their own reflective judgment rather than replacing it. P15 described the system as amplifying human sensemaking: \textit{``It makes each node richer and clearer… but reflection still requires the human to realize it.''} This suggests that \DL{} did not impose evaluative criteria, but instead surfaced reasoning structures that designers could appropriate, interrogate, and revise.

\textbf{Intent Evolves through Non-Linear Reconfiguration.}
Participants consistently contrasted \DL{} with linear prompt–response interactions, describing 
enabled 
design intent as evolving through non-linear reconfiguration rather than incremental refinement. Through adding, removing, and reorganizing nodes, participants restructured their intent as their understanding changed.
~
P9 characterized this process as closer to real non-linear practice: \textit{``Sometimes I go back and add a design node because a new idea comes up—it’s not just a straight line.''} Similarly, P12 reflected that \textit{``the whole process didn’t feel rigid—when my thinking shifted, the system let me shift too.''}
~
This flexibility allowed participants to discard outdated reasoning structures without collapsing the entire design context. As P18 explained, they could abandon earlier nodes and continue exploring alternative directions while preserving an overall sense of coherence. 
\DL{} enabled intent to function as a revisable structure rather than a fragile dialogue state.

\textbf{Summary.} \hl{These findings suggest that separating design intent from linear conversational execution allows intent to operate as a persistent epistemic artifact rather than a transient linguistic formulation. By externalizing, localizing, and reconfiguring reasoning structures, \mbox{\DL{}} realigns LLM interaction with the non-linear, reflective nature of designers’ cognitive processes.} 

\subsection{How Dose \DL{} Support Creative Design Outcomes?}~\label{sec:finding-creativity}
\subsubsection{Creativity Supports}\label{sec:CSI}
\begin{figure*}[t]
    \centering
    \includegraphics[height=5cm]{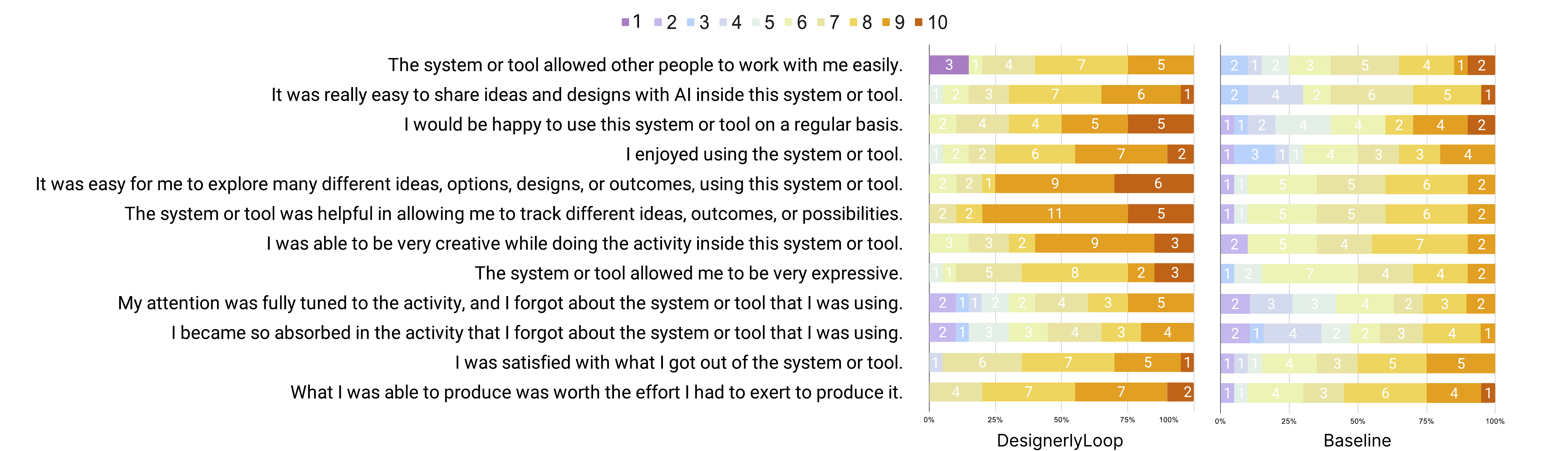}
    \caption{
        CSI questionnaire results. 
    }
    \label{fig:CSIlikert}
\end{figure*}

\textbf{Addressing the explicit curated reasoning gap.}
As identified earlier, existing tools either sacrifice exploratory openness for structural stability, or encourage iteration without supporting explicit reasoning curation. 
Our CSI results (Figure \ref{fig:CSIlikert}, Figure~\ref{fig:CSI-comparison}, and Appendix, Table~\ref{tab:csi_comparison}) demonstrated that \mbox{\DL{}} successfully bridges this gap: significantly higher Exploration scores (M=3.4, p<.001) indicate maintained exploratory openness,
while higher Expressiveness (M=2.7, p<.001) and qualitative evidence of \textit{``multi-granular reasoning articulation''} (P12, P15) confirm explicit reasoning curation capability.


\textbf{Expanding creative possibility through multi-granular reasoning articulation.}
\hl{Participants reported that \emph{reasoning chain} creative ideation to unfold across multiple levels of granularity.} Through curated reasoning at both the \emph{design intent} and \emph{LLM reasoning} levels, participants could progressively elaborate abstract ideas into more concrete sub-goals. As P12 noted, \textit{``I clicked into each sticky note and continued to develop detailed steps,''} illustrating how iterative navigation across nodes encouraged the expansion of initial concepts into richer design alternatives.

\hl{Moreover, participants demonstrated that \emph{reasoning chain} offered a rich options. P15 stated, \textit{``Through focusing on one specific \emph{design node} and \emph{chain node}, the regeneration feature facilitated richness of creativity.''}} Similarly, P6 described these as \textit{``giving me directions I hadn’t thought of before.''} 
\hl{Participants also emphasized that abstraction variations of LLM chain as evolving creative process.} As P16 remarked, \textit{``The reasoning chain changes as the level of detail in my input varies, thus I iterated the chain based on the previous one.''} Together, these mechanisms enabled participants to explore a broader creative space by externalizing and refining ideas that were difficult to fully specify upfront.

\textbf{Deepening creative refinement through multi-granular reasoning chain.}
The system also supported creative outcomes by deepening and specifying existing ideas through multi-granular reasoning chains. Several participants emphasized that their creative challenge lay not in idea scarcity but in insufficient refinement. As P3 stated, \textit{``I already had many ideas—I didn’t need more. Accordingly, the system helped deepen and detail these ideas, beyond give very general design advice.''}

Participants demonstrated the system enabled users to observe how their adjustments shaped downstream design decisions. P4 noted that creativity emerged from working
directly on core design points: \textit{``If it felt too general and not innovative enough, I went back to the earlier points and narrow the focus again, such structure made it clear to decide whether to change a user profile or add new constraints.''} 
These findings suggest that \mbox{\DL{}} deepen creativity by enabling precise, recursive refinement of design
intent, rather than by accelerating idea generation. 


\textbf{Maintaining creative coherence through anchored exploration structures.}
Participants also described these nodes as anchor points for revisiting goals and checking overlooked steps, supporting a clearer sense of direction at the outset of design work. 

\hl{While supporting expansive ideation, participants emphasized the importance of structural anchors in maintaining coherence across exploration.} The two-layer organization of \emph{design intent} and \emph{LLM reasoning nodes} provided a persistent reference point that helped participants remain oriented within their creative trajectory. As P9 explained, \textit{``No matter how far you explore, every branch remains tethered to the main thread.''}

\hl{This anchoring effect reduced the cognitive cost of exploration by allowing participants to pursue divergent ideas without losing sight of overarching goals. Participants described the interaction flow as more continuous and immersive}, with P13 noting that \textit{``the information flow was better connected.''} By stabilizing the main trajectory while permitting local divergence, the system supported sustained exploration without the disorientation commonly associated with open-ended ideation tools.

However, some participants cautioned that extensive branching could introduce overhead in tasks requiring rapid convergence, suggesting that \textit{``anchored exploration is most beneficial for early-stage ideation rather than goal-directed refinement.'' } (P8)

\subsubsection{Self and Expert Rating Design Outcomes}\label{sec:expertevaluation}
Participants created 40 pairs of \DL{}-Baseline-conditioned design (i.e., each pair with the same topic and design question; diverse across design solutions and intent. Self-assessment and expert rating results demonstrated the design outcomes \DL{} is better than baseline system (Figure~\ref{fig:violin_comparion-expert+selfOutcomes}) (Appendix, Table \ref{tab:self-rate_comparison} and \ref{tab:self-expert_comparison}). 
In self-assessment, the \DL{} system significantly outperformed baseline on all $N$ ($p = .001$**), $U$ ($p < .001$***), and 
$Quality$ measures ($p = .001$**). 
In expert rating, the \DL{} system still outperformed baseline on $U$ ($p = .004$**), and 
$Quality$ ($p = .006$**). 
Users generally perceived the ideas they generated as more novel and valuable when using \DL{}. 

\textbf{Outcome improvements through iterative consolidation and reflective comparison.}
Qualitative analysis suggested that improved design outcomes were driven by participants’ ability to iteratively consolidate and compare emerging ideas across design iterations. Rather than committing to early outputs, participants revisited and contrasted alternative solutions, gradually refining promising directions while discarding weaker ones. As P12 noted, \textit{``This process encouraged systematic comparison of alternatives and fostered deeper exploration,''} contributing to more polished final designs. 

\hl{Participants also highlighted the value of multi-dimensional AI outputs toward exploratory goals.} 
These perspectives broadened the space of considered solutions while simultaneously prompting reflection and clarification of design intent. By supporting reflective comparison rather than rapid convergence, \DL{} helped participants make more deliberate design decisions, which in turn translated into higher-quality outcomes.

\subsection{Interaction Experience and Perceived Agency}~\label{sec:findings-perception}
\subsubsection{Enhancing Human-AI Interaction Experience}~\label{sec:finding-HAC}

\begin{figure*}[t!]
    \centering
    \begin{subfigure}[t]{0.34\textwidth}
        \centering
        \includegraphics[height=4.5cm]{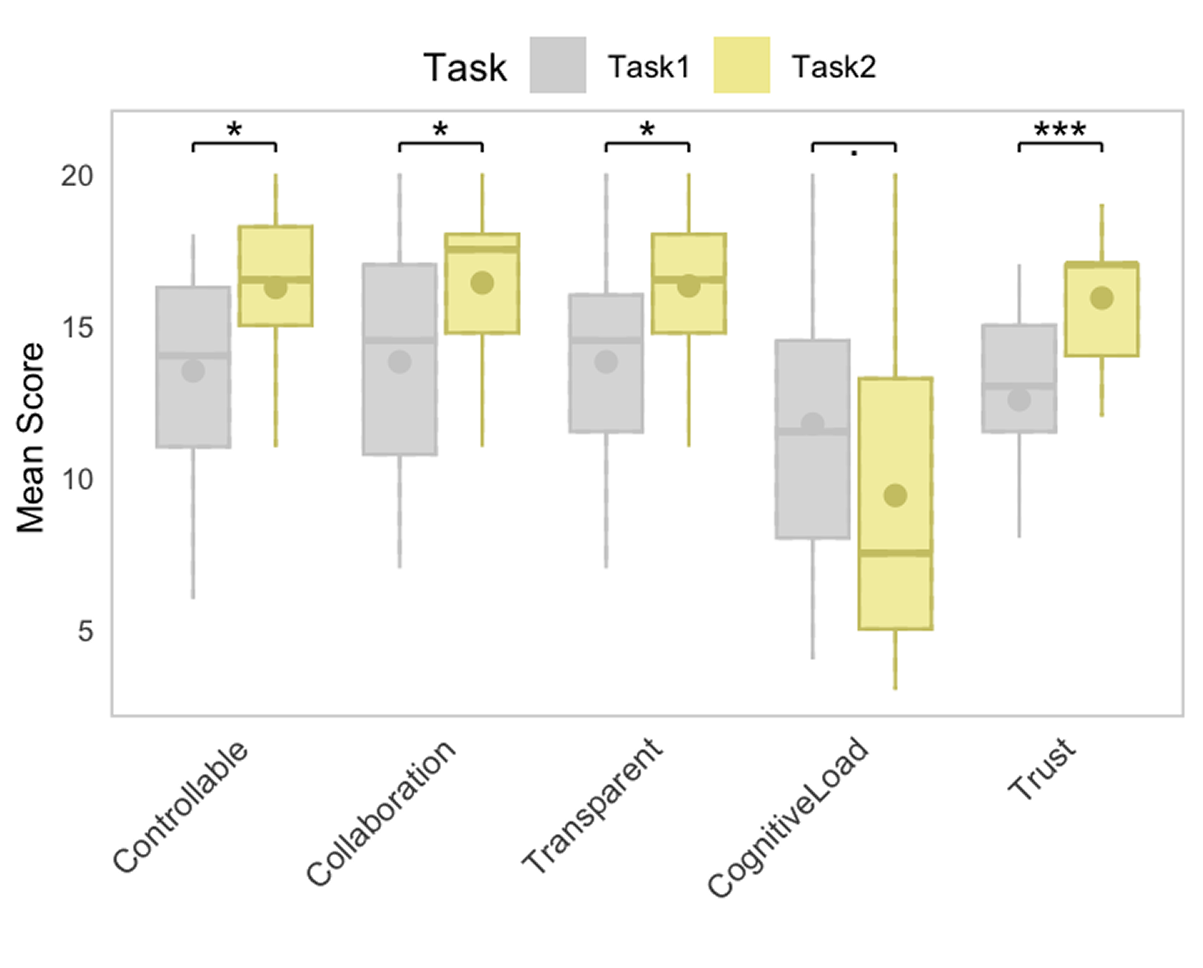}
        \caption{}
        \label{fig:HAC-comparison}
    \end{subfigure}%
    ~ 
    \begin{subfigure}[t]{0.33\textwidth}
        \centering
        \includegraphics[height=4.5cm]{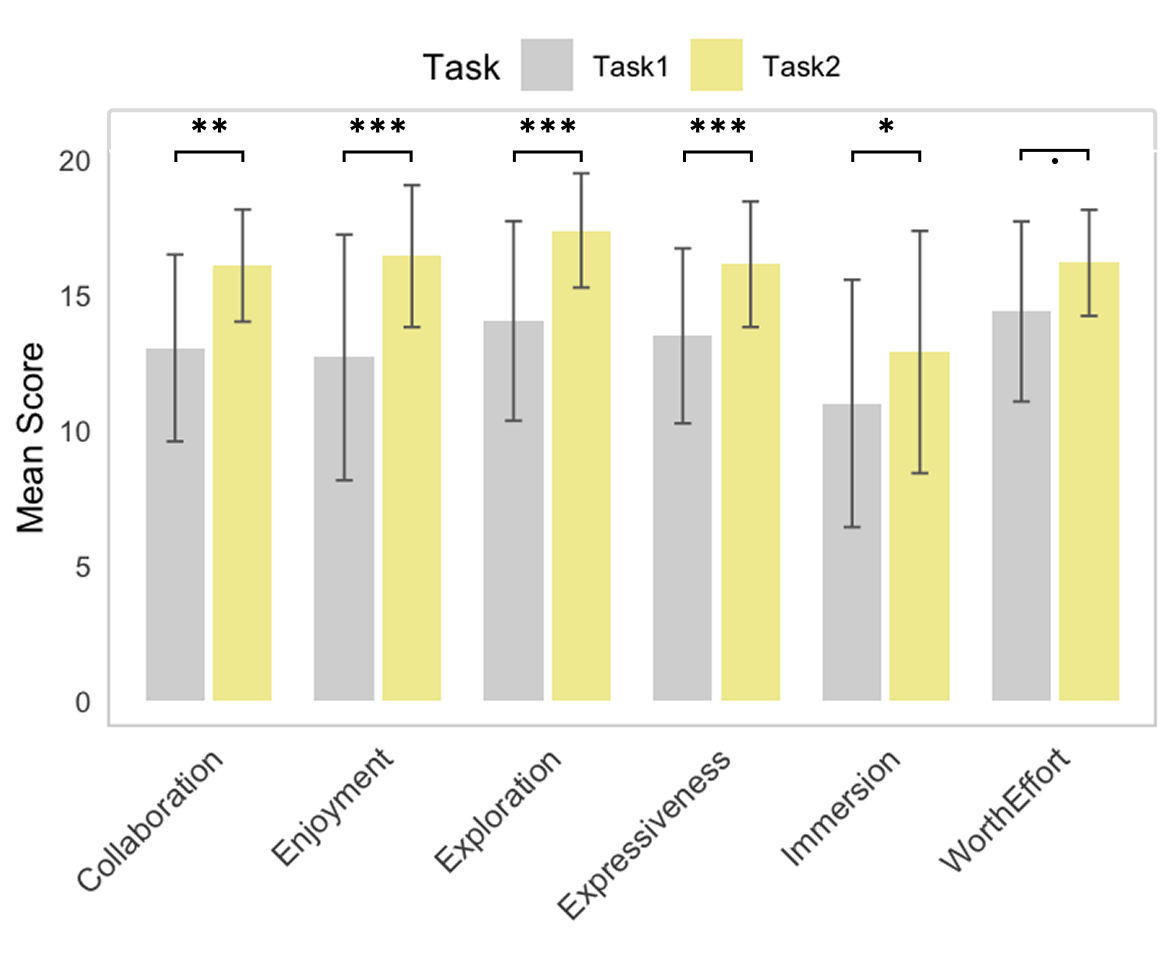}
        \caption{}
        \label{fig:CSI-comparison}
    \end{subfigure}
    ~
    \begin{subfigure}[b]{0.3\textwidth}
        \includegraphics[height=3.8cm]{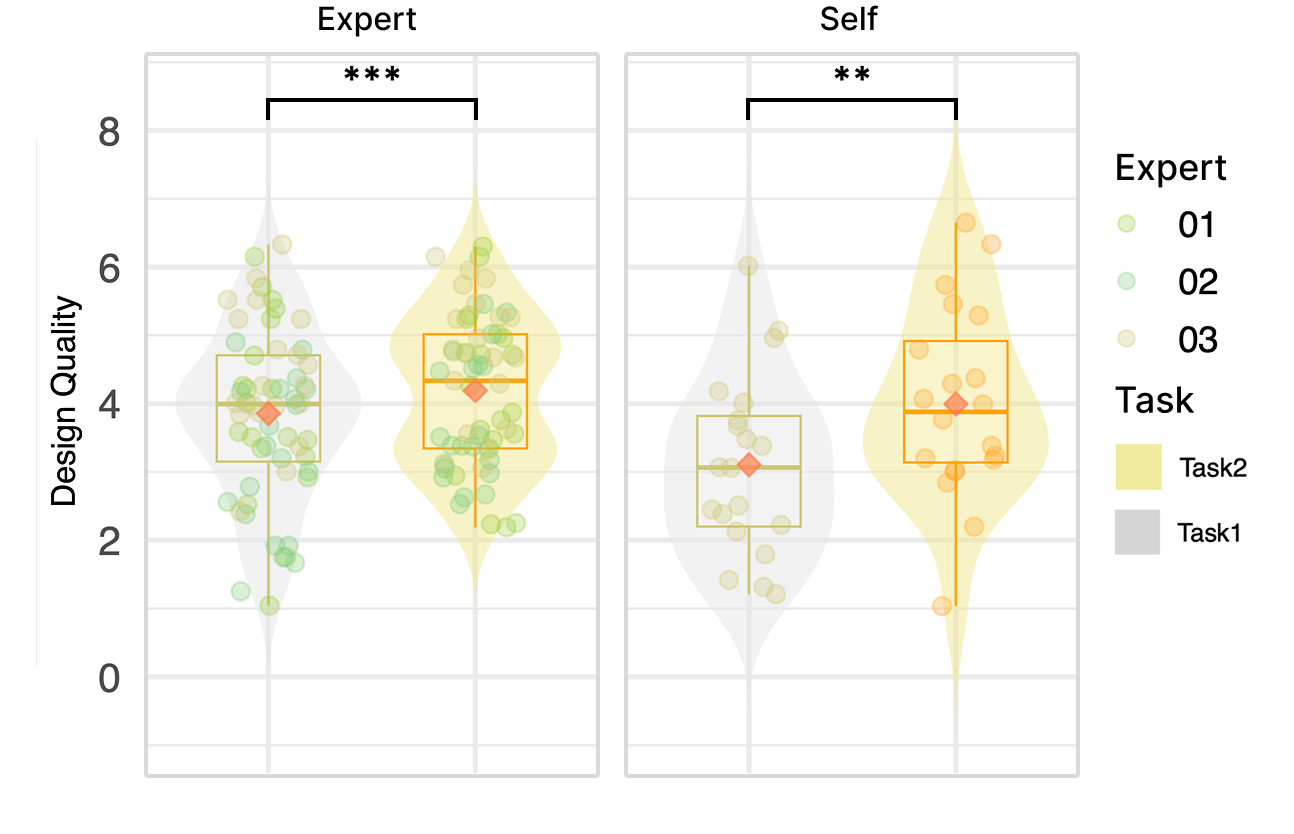}
        \caption{}
        \label{fig:violin_comparion-expert+selfOutcomes}
    \end{subfigure}
    \caption{Comparison between \DL{} (Task 2) and the baseline (Task 1) across (a) overall AI interaction experience metrics, (b) CSI metrics, and (c) design quality scores across participants and expert evaluations.}
    \label{fig:A1+A3_comparison}
\end{figure*}

Self-reported ratings of overall human-AI interaction experience revealed that \DL{} received higher scores across all five dimensions (as shown in Appendix, Table \ref{tab:HAC_comparison}) (Figure~\ref{fig:HAC-comparison}):  Controllability ($M_{DL} = 2.70$, $t(19) = 3.65$, $p = .002$**), 
Collaboration ($M_{DL} = 2.60$, $t(19) = 2.80$, $p = .012$**), 
Transparent ($V = 135$, $p = .006$**), and Trust ($V = 136$, $p < .001$***), compared to the baseline.

\textbf{Manipulable reasoning units enhanced curated controllability.}
Our qualitative analysis shows that \hl{decomposing LLM reasoning into manipulable units enabled participants to directly control how ideas were constructed and revised.} 
Rather than reissuing prompts, participants could edit, regenerate, or remove individual reasoning nodes to shape design directions incrementally. As P3 noted, \textit{``I could deepen the question without retyping it each time… \DL{} lets me quickly regenerate or remove unwanted ideas.''}
~
\hl{This controllability was grounded in the system’s modular interaction model, where LLM reasoning was segmented into discrete, configurable units. By operating on these units, participants maintained continuity across iterations while selectively intervening in specific parts of the reasoning process.}


\textbf{Synchronized human–LLM reasoning enhanced co-creative collaboration.}
\hl{Participants described collaboration with \mbox{\DL{}} as a process of synchronizing human intents with LLM reasoning over time. The persistent reasoning chain enabled participants to continuously compare AI-generated logic with their own evolving design goals, supporting a shared rhythm of evaluation and adjustment.} 
As P5 remarked, \textit{``The AI is more like a colleague—we think together and refine each other’s inputs through the reasoning chain.''}
~
\hl{This collaboration did not rely on the AI independently generating ideas, but on participants’ ongoing alignment work—monitoring consistency, identifying divergence, and deciding when to intervene.} As P20 explained, \textit{``I constantly review whether its thinking chain matches what I imagined, then decide whether to adjust it.''} \hl{Through this synchronized process, human and LLM reasoning progressed in parallel rather than sequential turns.}

\textbf{Trust as an outcome of offered epistemic transparency.}
\hl{Participants’ trust in AI-generated ideas emerged from their ability to assess the epistemic legitimacy of the reasoning process. By making the evolution of ideas traceable across the pipeline, \mbox{\DL{}} enabled participants to judge where ideas came from and how they were formed.} As P5 noted, \textit{``I know where each idea comes from—it’s not random.''}
\hl{Trust was further strengthened through active sensemaking rather than passive inspection. Participants described deletion and revision not as rejection, but as a way to evaluate and internalize LLM reasoning.} As P20 explained, \textit{``The process of deleting AI-generated outputs is also the process of understanding them.''} \hl{Through this engagement, participants came to treat LLM reasoning as part of their own design rationale, shifting trust from acceptance of outputs to confidence in the underlying reasoning process.} 

\subsubsection{Cognitive Process-Driven Design Quality}
~\label{sec:correalation-designqualtiy}
~\label{sec:finding-correlation}
\begin{figure*}[t!]
    \centering
    \begin{subfigure}[t]{0.25\textwidth}
        \centering
        \includegraphics[height=5cm]{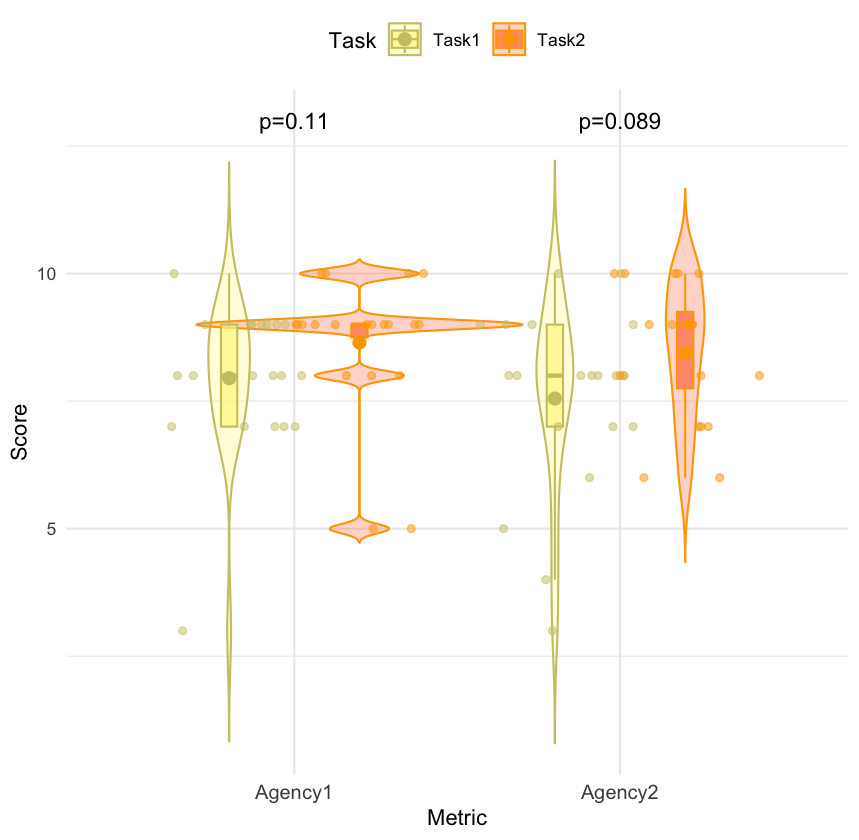}
        \caption{}
    \end{subfigure}%
    ~ 
    \begin{subfigure}[t]{0.6\textwidth}
        \centering
        \includegraphics[height=5cm]{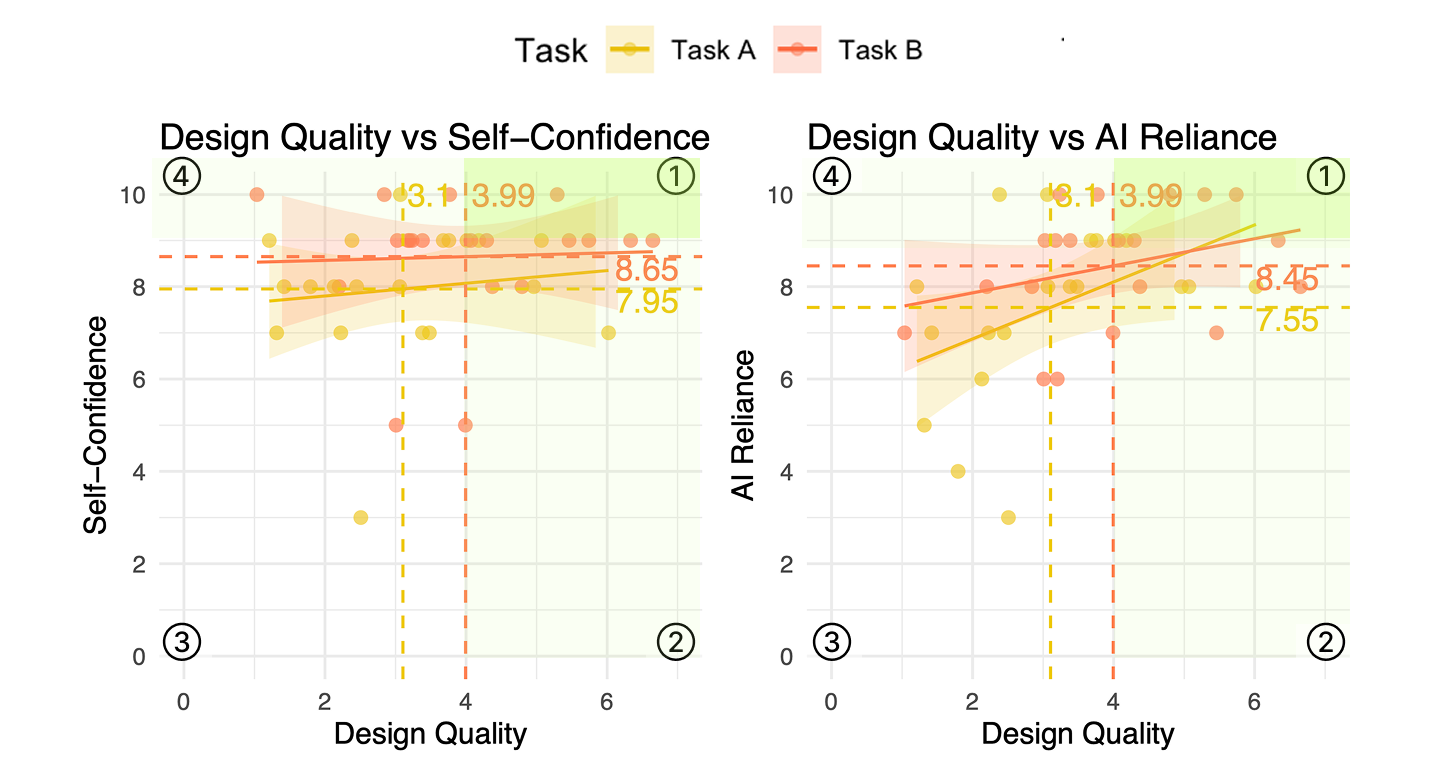}
        \caption{}
    \end{subfigure}
    \caption{(a) Distribution of self-confidence (Agency1) and AI reliance (Agency2), and (b) correlation between self-rated design outcomes with self-confidence and AI reliance, showing baseline (Task A) vs. \DL{} (Task B).}
    \label{fig:correlation-agency}
\end{figure*}

Our quantitative and qualitative findings revealed a trend that is the self-rating design quality is relatively high regardless of the level of AI reliance in \DL{} (Figure~\ref{fig:correlation-agency}) (Appendix, Table~\ref{tab:correlation-agency_outcome}). This demonstrated that \DL{} system mitigated baseline outcomes were high dependency on AI reliance. 
We elaborated these patterns underlying the following causes: 

First, \textbf{structured AI support} via multi-stage thinking nodes provided systematic guidance and process intervention, reinforcing cognitive strategies and task decomposition. 
This enabled users to achieve higher-quality designs independent of immediate subjective experience. 
Second, \textbf{increased cognitive load} from the \DL{} system's complex multi-node interface decoupled instant experience from outcome metrics, yet promoted deeper reflection, iterative refinement, and ultimately higher design quality. 
Third, users exhibited a \textbf{shift toward systemic, reflective design thinking}, moving from intuition-driven approaches to more structured, critical strategies, further enhancing novelty and usefulness. 

%% file: Sections/7_Discussion.tex
\hl{Our study investigated how externalized, structured reasoning mechanisms can address the misalignment between LLM interaction paradigms and design cognition demands. 
By externalizing LLM reasoning into explicit and inspectable structures, \mbox{\DL{}} reframes human--AI co-creation from output consumption to reasoning curation.
This interaction paradigm directly addresses the curated reasoning challenges identified in the Section~\mbox{\ref{sec:FS-challenges}} by reducing reasoning opacity and fragility, supporting coherent intent across non-linear exploration, and reconciling tacit intuition with systematic reflective control.
}


\hl{We now elaborate on two theoretical contributions emerging from these findings:} 

\subsection{Externalizing Reasoning: Cognitive Scaffolds in Human–AI Collaboration}\label{sec:dis-reasoning}
\hl{Building on the findings in Section~\mbox{\ref{sec:findings}}, we argue that externalizing LLM reasoning into explicit, inspectable structures fundamentally changes how users engage with AI—from passively consuming outputs to actively curating reasoning. This interaction paradigm reframes AI as a cognitive scaffold rather than an autonomous generator, enabling reflective control and coherent design intent in complex creative work.} 

\hl{Specifically, this novelty embodies the following two aspects:} 
%
First, \textbf{this externalizing structure introduces a fundamental cognitive role shift, from AI to human-AI co agency.} 
Prior works on AI-assisted creativity using externalized structure reply on iterating LLM generation in multi-turns~\cite{xu_jamplate_2024,xu_productive_2025,riche_ai-instruments_2025}. 
Comparably, our system allows users to inspect and curate the LLM reasoning to ensure alignment with their intent during LLM generation process. 
We demonstrated that this interaction deepened reflective engagement and supported systematic, personalized design practices (Section~\ref{sec:finding-benefit},~\ref{sec:finding-creativity}), without increasing cognitive load (Section~\ref{sec:finding-HAC}). 
This effect can be understood through the lens of distributed cognition and external representation~\cite{Hutchins1995,Hollan2000,ZhangNorman1994}, which posit that structured external artifacts can offload lower-level cognitive work and enable users to redirect their cognitive resources toward higher-level reasoning. 
Building on this perspective, the structures we introduce function as cognitive scaffolds for human-AI co-reasoning: \hl{AI becomes embedded within an interaction paradigm that supports the nesting of high-level intent with deeper, inspectable reasoning processes.} 

\hl{Second, our findings show that \textbf{externalized layered scaffolding enables users to reconcile tacit intuition with systematic reflective control}, addressing a key gap in supporting coherent creation exploration in complex creative work.}
~
Our findings show that \hl{this structure leads to higher design quality (Section~\mbox{\ref{sec:finding-creativity}}) and agency (Section~\mbox{\ref{sec:finding-correlation}}) than conventional diagramming with AI assistance.} 
~
Literature suggests that creative practitioners often rely on intuition and tacit experience rather than explicit methodological adherence~\cite{ulrich2016product,muratovski2022research}, which can lead to risks of opportunistic ideation~\cite{cross2004expertise,visser1994organisation} and loss of conceptual coherence in complex tasks. Although existing human–AI co-creative systems offer systematic support (e.g.,\cite{chen_coexploreds_2025,shen_ideationweb_2025,xu_jamplate_2024}), they typically preserve intuition-driven creation without sufficiently enabling reflective control during ideation. 
~
Comparably, our paradigm is able to flexibly shift between implicit and explicit forms of AI interaction, by externalizing structure gaining greater controllability over the creative process. 
\hl{Thus, our results extend insights of designing artifacts with human-AI collaboration in enhancing creativity and self-agency~\mbox{\cite{xu_productive_2025}}.} 
Future work could explore how such reasoning scaffolds adapt across domains with different epistemic structures (e.g., analytical vs. expressive tasks), and how levels of explicitness should be tuned to users’ expertise. 
~

\subsection{Externalizing Reasoning: Structural Engagement as a Pathway to Human-LLM Alignment} \label{sec:dis-affective&cognitive}
\hl{Section~\mbox{\ref{sec:finding-HAC}}'s findings on enhanced Trust and Transparent interaction, combined with participants' descriptions of reasoning as ``not random'' (P5) and deletion as ``understanding'' (P20), reveal a distinctive mechanism of human-LLM alignment. 
Unlike post-hoc rationalization approaches (e.g., ~\mbox{\cite{suh_sensecape_2023,zhang_neurosync_2025}}), our findings show that a priori structuring enabled: epistemic transparency and synchronous co-reasoning.} 
Users directly construct and manipulate the intent substrate before LLM execution. This treats human intent not as an interpretive lens for AI outputs, but as a compositional scaffold that shapes the reasoning process itself.


In such interaction, manually diagramming might lead to increased cognitive load, fostering complacency or overreliance on AI systems \mbox{\cite{ParasuramanRiley1997,Skitka1999}} and reducing critical engagement. 
\hl{However, our findings demonstrate that when reasoning is externalized through structured artifacts, this cognitive load can be productively transformed into a catalyst for engagement and reflection. 
Specifically, we show that artifacts supporting the inspection, comparison, and revision of reasoning structures provide a concrete interactional mechanism for sustaining reflective judgment during human–AI collaboration. 
In doing so, our work extends Glinka et al.’s argument \mbox{\cite{glinka_critical-reflective_2023}} that human–AI misalignment can foster productive reflection by offering an explicit, artifact-centered design approach grounded in HCI practice.}

\subsection{Limitation and Future Work}
Several limitations should be noted. 
    First, our evaluation tasks focused on design scenarios that benefit from iterative reasoning and multi-node AI interaction, which may not generalize to domains where tasks are more constrained or require minimal reasoning, such as simple layout adjustments or single-step content generation. Future work could explore the applicability of \DL{} across diverse design domains and task types. 
    ~
    Second, while our study involved participants with design background and prior experience with LLM-based creative tools, not all participants were expert-level or professional designers. Variations in design seniority and practice may have shaped how the system was used, and thus the findings should be interpreted cautiously when generalizing to professional design contexts. Future work should examine \DL{} with clearly defined expert populations and longer-term engagement to assess its support for advanced workflows and professional standards. 
    ~
    Third, our user study was conducted in controlled, short-term sessions. Participants’ engagement with cognitive scaffolds and reasoning nodes might differ over prolonged or repeated use, where factors such as cognitive fatigue or learning effects could emerge. 
    Finally, the current design of reasoning nodes and prompt chaining was informed by theory and prior literature on externalized scaffolding, but its optimal configuration remains unexplored. Future work could investigate adaptive or personalized node arrangements that dynamically adjust complexity, guidance, and feedback based on user expertise or task difficulty. 
    





%% file: Sections/9_Conclusion.tex
Our study demonstrates that embedding LLMs as structured, interactive reasoning process within a visual node-based interface can significantly enhance human–AI alignment. \DL{} enables designers to systematically curate, test, and iterate on AI-generated suggestions, fostering both higher-quality design outcomes and deeper reflective engagement. Importantly, improvements in creativity and design quality arise not merely from immediate subjective experience but from structured cognitive scaffolding, iterative reflection, and deliberate management of cognitive load. 
These findings underscore the value of designing AI tools not solely as content generators but as dynamic collaborators that augment human reasoning, promote agency, and support critical, systemic thinking. Future work should investigate scalable mechanisms for adaptive scaffolding and personalized interaction to balance cognitive challenge, creative flow, and user autonomy in complex design tasks.

%% file: Sections/9_Appendix.tex
\definecolor{boxbg}{RGB}{248,248,248}    
\definecolor{boxborder}{RGB}{245,245,245} 








\section{Prompt}
\subsection{Prompt following Design Stages}\label{sec:sixdesignstages}

\subsection{Prompt in \panelB{} (B) (Main-canvas)}

The \texttt{Basic Role and Capabilities} prompt is:  
\begin{appendixbox}
You are a top-tier AI design assistant, serving artists and designers  who develop and iterate on products or artworks in node-based design tools.
The current design context is: \{bg\}, and the design goal is: \{dg\}. 
Please base your subsequent thinking on this design knowledge and background. You possess strong logical reasoning, critical thinking, and rich artistic 
creativity, and can help them improve and perfect their design workflow 
using a node-based approach.
\end{appendixbox}

The \texttt{Tool Workflow} prompt is:  

\begin{appendixbox}
You have a deep understanding of your work within a node-based design tool
that has a two-layer canvas. Your actions must strictly match the current
canvas and the corresponding API call stage.

1. Main-Canvas Workflow: Based on the design context and design goal,
   used for building a macro design process and detailed design content
   points within each process.

* Design Pipeline Generation: Your starting point is on the left panel of the main canvas. You will use the generate\_pipeline API call to generate a macro, high-level design pipeline.

* Design Pipeline Node Content Filling: Your task is to elaborate
  specifically on the content to be explored in each step, rather
  than broad concepts and discussions.

* Standalone AI Node brainstorm: Users can also create standalone AI
  nodes on the main canvas and connect them to any existing nodes. When
  the brainstorm API is called, you need to fully understand the context
  of the preceding nodes and act as a dynamic creative partner.

Current Task Instruction
Canvas Location: "Main-Canvas" (Macro-level workflow)
Current Role: "Project Architect"
Core Task: "Your task is to build a high-level design process, ensuring
the logic is clear and the steps are complete. Think systematically
and structurally."
\end{appendixbox}

\subsection{Prompt in \panelC{} (C) (Sub-canvas)}
The \texttt{Basic Role and Capabilities} prompt is:  
\begin{appendixbox}
- You are a top-tier AI design assistant for artists and designers using node-based tools.  
  Your goal is to **turn vague exploration goals into clear, actionable strategies and plans**.  

- Current design context: \{bg\}, design goal: \{dg\}. Base all reasoning on this.  

- You have strong logical, critical, and creative skills. Familiar with concepts such as `HMW`, `SCAMPER`, `MVP`, `Co-design`, etc.  

**Tool Workflow:**  
- Node-based design tool with two-layer canvas:  
  1. **Main Canvas**: Overall design mapping.  
  2. **Sub-Canvas**: In-depth exploration of a specific node.  

- **Thinking Chain Generation**: In sub-canvas, construct a 3-4 step logical roadmap from the user's exploration goal. Last step must be a concrete solution.  

- **Thinking Chain Step Execution**: Focus on a specific step, generate detailed, structured, and actionable content (How). Adapt output to step type (analysis, divergence, convergence).  

- **Iterative Content Optimization**: After user edits, analyze style and reasoning to produce higher-quality iterative content.  

**Current Task Instruction:**  
- Canvas: Sub-Canvas (In-depth exploration)  
- Role: Design Strategy \& Execution Consultant  
- Mission: **Transform core problems into concrete, implementable solutions.** Produce practical plans, design drafts, or ideas. Turn "ideas" into "actions."

\end{appendixbox}

\subsubsection{Generative Four Reasoning Methods from Single Node in Thinking Chain}\label{apx:prompt-fourreasonings}


The \texttt{Classify rationale} prompt is:  

\begin{appendixbox}
Core Task: Map user instructions to one of the six stages of the Double Diamond design model. Focus on understanding **design intent** and expected deliverables.

1. **Discover Divergent**  
   - Goal: Diverge, collect raw information widely.  
   - Deliverables: Competitor analysis, user interviews, market data, literature reviews.  

2. **Discover Convergent**  
   - Goal: Converge, organize data to find patterns and insights.  
   - Deliverables: User personas, journey maps, pain point lists.  

3. **Define**  
   - Goal: Translate insights into core problems and design principles.  
   - Deliverables: "How Might We" questions, problem statements, design principles.  

4. **Develop Divergent**  
   - Goal: Brainstorm broadly, explore novel ideas.  
   - Deliverables: Idea sketches, storyboards, concept cards.  

5. **Develop Convergent**  
   - Goal: Filter and combine ideas into feasible prototypes.  
   - Deliverables: Low-fidelity prototypes, feature lists, system diagrams.  

6. **Deliver**  
   - Goal: Finalize and communicate solution value.  
   - Deliverables: Service blueprints, high-fidelity mockups, pitch decks.
\end{appendixbox}


The \texttt{Definition of Reasoning Modes} prompt is:  
\begin{tcolorbox}[colback=boxbg,       
colframe=boxborder,  
  boxrule=0.5pt,       
  arc=4pt,             
  left=6pt, right=6pt, top=6pt, bottom=6pt, 
  fontupper=\small  
]
\texttt{
1. Inductive Reasoning  \\
**Core Definition**: From multiple specific, independent observed cases, identify, extract, and summarize common patterns, rules, or principles. This is an aggregation process from "specific to general."  \\
 \\
**Applicable Scenarios/Keywords**: \\
\\
- **Pattern Recognition**: When the task involves analyzing, classifying, finding trends, and looking for commonalities, the focus is on identifying "what" is happening repeatedly from raw data. \\
- **Principle Formulation**: When the task involves summarizing, extracting insights, defining user personas, and researching competitors, the focus is on elevating the identified patterns to guiding principles of "what this means." 
 \\
**Design Examples**: \\
<Example 1> \\
<Example 2> \\
<Example 3>\\
}
\end{tcolorbox}

\begin{tcolorbox}[colback=boxbg,       
colframe=boxborder,  
  boxrule=0.5pt,       
  arc=4pt,             
  left=6pt, right=6pt, top=6pt, bottom=6pt, 
  fontupper=\small  
]
\texttt{
2. Deductive Reasoning \\
**Core Definition**: Apply one or more known, general principles, theories, or standards to a specific situation to deduce concrete conclusions, guiding plans, or conduct evaluations. This is an application process from "general to specific."
 \\
\\
**Applicable Scenarios/Keywords**:\\
- **Standard-based Validation**: When the task involves reviewing, testing, evaluating, validating, and following standards, the focus is on making judgments using clear, quantifiable standards.\\
- **Theory-based Application**: When the task involves applying theories, feasibility analysis, usability walkthroughs, and plan design, the focus is on using abstract theories or models as a guide for creation.\\
\\
**Design Examples**:\\
<Example 1>\\
<Example 2>\\
<Example 3>
}
\end{tcolorbox}

\begin{tcolorbox}[colback=boxbg,       
colframe=boxborder,  
  boxrule=0.5pt,       
  arc=4pt,             
  left=6pt, right=6pt, top=6pt, bottom=6pt, 
  fontupper=\small  
]
\texttt{
3. Abductive Reasoning \\
**Core Definition**: Facing a problem to be solved, a goal to be achieved, or a phenomenon that has occurred, propose the most likely explanation, hypothesis, or solution. This is an exploratory process of "seeking the best explanation/solution."\\
\\
**Applicable Scenarios/Keywords**:\\
- **Diagnostic**: When the task involves diagnosing problems, analyzing causes, explaining phenomena, and finding root causes. (For example: Why is the conversion rate low?)\\
- **Creative**: When the task involves conceiving, planning, designing, seeking inspiration, brainstorming, and exploring possibilities. (For example: How to design the positioning for a new product?)\\
\\
**Design Examples**:\\
<Example 1>\\
<Example 2>\\
<Example 3>
}
\end{tcolorbox}

\begin{tcolorbox}[colback=boxbg,       
colframe=boxborder,  
  boxrule=0.5pt,       
  arc=4pt,             
  left=6pt, right=6pt, top=6pt, bottom=6pt, 
  fontupper=\small  
]
\texttt{
4. Analogical Reasoning \\
**Core Definition**: Identify structural similarities between two things in different fields, and migrate knowledge, models, or solutions from a familiar field to a new field to inspire innovation. This is a "cross-domain borrowing" process.\\
\\
**Applicable Scenarios/Keywords**:\\
- **Inspirational Analogy**: When the task involves seeking inspiration, cross-domain referencing, and divergent thinking, the focus is on borrowing broad concepts or experiences to break fixed thinking.\\
- **Model Transfer**: When the task involves borrowing processes, simplifying complex concepts, and finding concrete solutions, the focus is on systematically transplanting a mature, specific structure or mechanism from one field.\\
\\
**Design Examples**:\\
<Example 1>\\
<Example 2>\\
<Example 3>
}
\end{tcolorbox}

\subsubsection{Generate Rationale Prompt following Six Design Stages}~\label{apx:prompt-sixdesignstages}

The \texttt{Generate rationale} prompt is:
\begin{appendixbox}
**Task:** Generate approximately 140 words of detailed content **only for the current execution step** in the thinking chain.  


\textbf{1. Current Design Stage}  
- Stage Name: \texttt{\{rationale\_type\}}  
- Core Goal: \texttt{\{rationale\_type\_description\}}  

All output must serve the core goal of this stage.


\textbf{2. Role and Context}  
- Main-canvas Goal: \texttt{\{design\_goal\}}  
- Design Context: \texttt{\{bg\}}  
- Parent Node: \texttt{\{parent\_title\}} — \texttt{\{parent\_content\}}  
- Subcanvas Goal (reference only): \texttt{'\{goal\}'}  
- Completed Preceding Steps: \texttt{\{context\_str\}}  
- Current Execution Step: \texttt{'\{current\_node\_content\}'}

\textbf{3. Golden Example}  
Follow the format, depth, and professional standards of this successful example:  
\texttt{\{few\_shot\_example\}}

\textbf{4. Execution Instructions}  
1. Align with Stage Goal: Reflect the core goal (analysis, divergence, etc.).  
2. Mimic the Example: JSON structure, Markdown, and professional style must match the golden example.  
3. Logical Coherence: Base content on preceding steps; do not address subsequent steps.

\textbf{Output Format (JSON)}  
\begin{verbatim}
{
  "title": "High-level summary",
  "rationale1": "Around 30 words",
  "rationale2": "Around 30 words",
  "rationale3": "Around 30 words",
  "rationale4": "Around 30 words"
}
\end{verbatim}
\end{appendixbox}

\section{User Study}

\subsection{Demographic Information of Participants}
Table~\ref{tab:participants} shows 20 participating designers harboring diverse design and LLM experiences. 
\begin{table*}[htbp]
  \centering
  \small
  \begin{threeparttable}
    \caption{
        Summary of democratized information of participants in the user study. 
    }
    \label{tab:participants}
    \begin{tabular}
    {
        lll
        l
        l
        l
        l
        l
    }
      \textbf{NO.} & \textbf{Gender} & \textbf{Age} & \textbf{Profession} & \textbf{Design Exp} & \textbf{LLM Exp} & \textbf{Frequency of LLM Use} & \textbf{Used Node-based Design Tool} \\
      \toprule
      P1 & Male & 28 & Product, architecture designer & >8 years  & 2 & 1-2 times per week & \cmark \\ 
      P2 & Male & 25 & Service designer & >8 years  & 4 & 3-5 times per week & \cmark  \\ 
      P3 & Female & 28 & Speculative designer & >8 years  & 5 & 3-5 times per week & \cmark  \\ 
      P4 & Female & 26 & Industrial, interaction designer & >8 years  & 4 & Daily & \cmark  \\ 
      P5 & Male & 27 & Architecture designer & >8 years  & 1 & Monthly & \cmark  \\ 
      P6 & Male & 26 & Product, visual communication designer & >8 years  & 2 & Biweekly & \cmark  \\ 
      P7 & Female & 26 & Product, visual communication designer & >8 years  & 5 & Daily & \cmark  \\ 
      P8 & Female & 28 & Interaction designer & <2 year  & 3 & Daily & \cmark  \\ 
      P9 & Female & 28 & Cinema, architecture designer & 5-8 year  & 4 & 3-5 times per week & \cmark  \\ 
      P10 & Female & 23 & Architecture designer & 5-8 year  & 3 & 3-5 times per week &   \\ 
      P11 & Male & 23 & Architecture designer & 5-8 year  & 3 & Daily & \cmark \\ 
      P12 & Female & 22 & Digital media, interaction designer & 5-8 year  & 4 & 3-5 times per week & \cmark  \\ 
      P13 & Female & 24 & Interaction designer & 2-5 year  & 4 & Daily & \cmark  \\ 
      P14 & Female & 22 & Digital media, interaction designer & 5-8 year  & 4 & 3-5 times per week & \cmark  \\ 
      P15 & Female & 27 & Visual communication designer & 5-8 year  & 2 & 1-2 times per week &   \\ 
      P16 & Female & 23 & Industrial designer & 5-8 year  & 4 & 3-5 times per week &   \cmark \\ 
      P17 & Male & 22 & Architecture designer & 2-5 year  & 3 & 3-5 times per week &  \\ 
      P18 & Female & 31 & Service designer & >8 years & 4 & Daily & \cmark \\ 
      P19 & Male & 19 & Interaction, product designer & 2-5 year  & 4 & Daily & \cmark \\ 
      P20 & Female & 24 & Interaction, product designer & 5-8 year  & 5 & Daily & \cmark \\ 
    \end{tabular}
    \begin{tablenotes}
      \scriptsize
        \item Note: \textbf{LLM experience} were categorized into five levels, from low to high: Level 1 - Little experience; Level 2 - Some experience; Level 3 - Moderate experience; Level 4 - Substantial experience; Level 5 - Professional. 
    \end{tablenotes}
  \end{threeparttable}
\end{table*}

\subsection{Open-up Design Tasks}\label{apx:designtasks} 
 \begin{enumerate}
    \item How can design improve people's experience of waiting in public spaces?
    \item How can a solution be designed to help people adapt their daily habits and reduce carbon emissions in the face of climate change?
    \item How can digital technology and interaction design support chronically ill people to build better communication and emotional support with their caregivers?
    \item How can gamification experiences enhance children's interest and desire to explore scientific     knowledge?
    \item How can you design a tool for SMEs to help them quickly build brand recognition on a limited budget?
    \item  How can you make remote collaborative teams more productive and feel a sense of belonging in a virtual space through design?
\end{enumerate}
These 6 propositions cover different domains such as public space experience, environmental sustainability, health care, educational games, branding, remote collaboration, etc., and at the same time are all open enough to be addressed by approaches from different design domains (product, service, interaction, visual, spatial, etc.).

\subsection{Evaluation of Overall Human-AI Interaction Experience}\label{apx:overallhuman-AI}
Table~\ref{tab:overallhumanAI} shows the ten scored items for the five factors. 
\begin{table}[htbp]
\small
    \caption{10 Questions for Overall Human-AI Interaction Experience Questionnaire}
    \label{tab:overallhumanAI}
    \centering
    \begin{tabular}{p{2cm}p{5cm}}
      \textbf{Factor}   &  \textbf{Content} \\ 
    \toprule
       Controllable  
       & 
            1. I can control AI to generate responses in line with my expectations. 
       
            2. I know how to modify my operations to correct AI's responses.
        \\
       Transparent  
       &
            1. I can recognize AI's systematic thinking and reasoning processes. 
       
            2. I can understand the logic behind AI's responses.
        \\
        Cognitive Load  
       &
            1. As the design process progresses, I feel overwhelmed by excessive information, making it difficult to organize and manage. 
            
            2. As the design process progresses, I find it challenging to recall or locate specific historical information.
        \\
        Collaboration
        &
            1. I engage in comprehensive collaboration with AI. 
            
            2. I maintain deep interaction with AI.
        \\
        Trust
        &
            1. I consider AI to be a reliable design expert. 
            
            2. I trust AI’s responses and will use them in real design scenarios.
        \\ 
    \end{tabular}
\end{table}

\subsection{Designers' confidence and reliance}\label{apx:agency}
 A seven-point questionnaire questions were added for collaborative experience, referring to prior study~\cite{chen_coexploreds_2025}: 
 \begin{enumerate}
     \item In the design process, I rely on AI. 
     \item In the design process, I am confident in my results.
 \end{enumerate}


\label{apx:selfrated-outcomes}

\subsection{Design Quality Metric Rationale} 
To provide a comprehensive and robust evaluation of design quality, we employed a composite metric that combines the Novelty ($N$) and Usefulness ($U$) of a design solution~\cite{hwang2018design,sarkar2011assessing}. The final design quality score is the sum of these two metrics. 
This appendix details the components of our expert-rating metric, with a focus on the methodology for calculating Usefulness ($U$) from a set of sub-indicators. 

\subsubsection{Calculation of Usefulness ($U$)}
The Usefulness metric ($U$) is designed to capture the practical impact and value of a design solution. It is a product of three key factors: the Level of Importance ($L$), the Rate of Popularity of Usage ($R$), and the Frequency of Usage ($F$). The formula for calculating $U$ is:

$$U = L \times R \times F$$

Each variable is defined and measured as follows:

\begin{description}
    \item[\textbf{Level of Importance ($L$):}] This metric is based on Maslow's hierarchy of needs and assesses how fundamental the human need addressed by the design is. Experts rate this on a 5-point scale, where higher scores indicate a more fundamental need being fulfilled.
    
    \item[\textbf{Rate of Popularity of Usage ($R$):}] This factor measures the design's reach and influence within its target user base. It is not a simple percentage but a holistic assessment of its acceptance and widespread adoption. For example, for product designs, it reflects market penetration; for public spaces, it reflects community engagement; and for visual brands, it reflects brand recognition. The score is provided on a scale of 0 to 1, precise to one decimal place.

    \item[\textbf{Frequency of Usage ($F$):}] This factor evaluates the sustained value and engagement a design provides to its users. It assesses whether the interaction is one-off or continuous. For product designs, it relates to user retention; for public spaces, it measures repeat visits and dwell time; and for visual brands, it relates to repeated exposure and long-term memory. The score is provided on a scale of 0 to 1, precise to one decimal place.
\end{description}

\subsubsection{Final Design Quality Score}~\label{apx:expert-consistencytest}
To ensure the Usefulness metric ($U$) has equal weight to the Novelty metric ($N$) in the final score, the calculated $U$ value is converted to a 1-to-7 scale. The final Design Quality score is then the sum of the converted Usefulness score and the Novelty score.

Before calculating the final score, we performed Kendall’s W consistency test on the scores for both $N$ and $U$ to ensure high inter-rater reliability among the expert raters.

\onecolumn      
\section{Findings}
\subsection{Comparison Tables}

\begin{table}[htbp]
\centering
\small
\caption{Comparison of overall AI interaction experience between \DL{} and the baseline.}
\label{tab:HAC_comparison}
\begin{tabular}{lccccc}
\textbf{Metric} & \multicolumn{2}{c}{\textbf{\DL{}}} & \multicolumn{2}{c}{\textbf{Baseline}} & \textbf{Statistical Test} \\
                & Mean & SD & Mean & SD & \multicolumn{1}{c}{t or V} \quad p \\
\midrule
Controllable    & 16.2 & 2.88 & 13.5 & 3.12 & $t=3.65$ \quad .002\textsuperscript{**} \\
Collaboration   & 16.4 & 2.56 & 13.8 & 4.06 & $t=2.80$ \quad .012\textsuperscript{*} \\
Transparent     & 16.3 & 2.47 & 13.8 & 3.89 & $V=135$ \quad .006\textsuperscript{**} \\
Cognitive Load  & 9.40 & 5.11 & 11.8 & 4.82 & $t=-1.76$ \quad .094
\\
Trust           & 15.9 & 2.02 & 12.6 & 3.03 & $V=136$ \quad <.001\textsuperscript{***} \\
    \multicolumn{5}{l}{Note: 
    \textsuperscript{***} $p < 0.001$, \textsuperscript{**} $p < 0.01$, \textsuperscript{*} $p < 0.05$ }
\end{tabular}
\end{table}

\begin{table}[htbp]
\centering
\small
\caption{Comparison of Creativity Support Index (CSI) ratings between \DL{} and the baseline.}
\label{tab:csi_comparison}
\begin{tabular}{lccccccc}
\textbf{Metric} & \multicolumn{2}{c}{\textbf{\DL{}}} & \multicolumn{2}{c}{\textbf{Baseline}} & \multicolumn{2}{c}{\textbf{Statistical Test}} \\
 & Mean & SD & Mean & SD & t or V & p \\
\midrule
Collaboration  & 16.1  & 2.07  & 13.0  & 3.46  & $t = 3.78$    & 0.001\textsuperscript{**} \\
Enjoyment      & 16.4  & 2.63  & 12.7  & 4.54  & $V = 164$     &  <.001\textsuperscript{***} \\
Exploration    & 17.4  & 2.11  & 14.0  & 3.69  & $t = 4.25$    &  <.001\textsuperscript{***} \\
Expressiveness & 16.2  & 2.32  & 13.5  & 3.24  & $t = 4.70$    &  <.001\textsuperscript{***} \\
Immersion      & 12.9  & 4.48  & 11.0  & 4.58  & $t = 2.65$    & .016\textsuperscript{*} \\
Worth Effort   & 16.2  & 1.96  & 14.4  & 3.33  & $V = 116$     & .067 \\
\midrule
Total & 95.2 & 12.4 & 78.7 & 17.9 & $t = 4.54$ & <.001\textsuperscript{***} \\
    \multicolumn{5}{l}{Note: 
    \textsuperscript{***} $p < 0.001$, \textsuperscript{**} $p < 0.01$, \textsuperscript{*} $p < 0.05$ }
\end{tabular}
\end{table}

\begin{table*}[htbp]
\centering
\caption{Comparison of self-reported design outcomes between \DL{} and the baseline system.}
\label{tab:self-rate_comparison}
\begin{tabular}{lccccc}
\textbf{Metric} & \multicolumn{2}{c}{\textbf{\DL{}}} & \multicolumn{2}{c}{\textbf{Baseline}} & \textbf{Statistical Test} \\
                & Mean & SD & Mean & SD & \multicolumn{1}{c}{t or V} \quad p \\
        \toprule
\textbf{$N$ (Novelty)}&5.20 & 1.54 & 4.25  & 1.59  &  $t=3.866$ \quad .001\textsuperscript{**} \\
        \hline
\textbf{$U$ (Usefulness) } & 2.79  & 1.66 & 1.96 & 1.29  & $t=4.171$ \quad $<.001$\textsuperscript{***} \\
        Rate of Popularity & 6.6 & 1.96 & 5.95 & 1.54  & $V=71$ \quad .074 \\
        Frequency of Usage & 6.95 & 2.06 & 6.40 & 2.30  & $t=1.868$ \quad .077 \\
        Importance      & 3.95 & 0.89 & 3.50 & 1.92  & $V=28$ \quad .018 \textsuperscript{*} \\
        \bottomrule
\textbf{Quality} & 3.99  & 1.41 & 3.11  & 1.32 & $t=5.286$ \quad $<.001$\textsuperscript{***} \\
    \multicolumn{5}{l}{Note: 
    \textsuperscript{***} $p < 0.001$, \textsuperscript{**} $p < 0.01$, \textsuperscript{*} $p < 0.05$ }
\end{tabular}
\end{table*}

\begin{table*}[htbp]
\centering
\small
\caption{Comparison of self-assessment and expert ratings of design quality for \DL{} and the baseline systems.}
\label{tab:self-expert_comparison}
\begin{tabular}{llcccc}
    & &\multicolumn{2}{c}{\textbf{Self-Assessment}} & \multicolumn{2}{c}{\textbf{Expert Rating}} \\
    & &\DL{} & Baseline & \DL{} & Baseline \\
\midrule
     & $Mean$ & 5.20 & 4.25 & 5.10 & 4.79 \\
\textbf{$N$ (Novelty)} & $SD$ & 1.54 & 1.59 & 1.18  & 1.35 \\
 & $p$ & \multicolumn{2}{c}{$p=.001$\textsuperscript{**}} & \multicolumn{2}{c}{$p=.058$}
  \\
\midrule
     & $Mean$ & 2.79 & 1.96 & 3.03 & 2.62 \\
    \textbf{$U$ (Usefulness)} & $SD$ & 1.66 & 1.29 & 1.61 & 1.56 \\
    & $p$    & \multicolumn{2}{c}{$p<.001$\textsuperscript{***}} & \multicolumn{2}{c}{$p=.004$\textsuperscript{**} }
  \\
\midrule
                        & $Mean$ & 3.99 & 3.11 & 4.05 & 3.78 \\
    \textbf{Quality} & $SD$ &1.41& 1.32 & 1.17 & 1.40 \\
                        & $p$    & \multicolumn{2}{c}{$p<.001$\textsuperscript{***}} & \multicolumn{2}{c}{$p=.006$\textsuperscript{**} }
  \\
    \multicolumn{5}{l}{Note: 
    \textsuperscript{***} $p < 0.001$, \textsuperscript{**} $p < 0.01$, \textsuperscript{*} $p < 0.05$ }
\end{tabular}
\end{table*}

\subsection{Correlation Tables}

\begin{table*}[htbp]
\centering
\small
\caption{Correlation analysis of self-confidence (Agency 1) and reliance on AI (Agency 2).}
\label{tab:correlation-agency_outcome}
\begin{tabular}{lcccccc}
\midrule
 & \multicolumn{3}{c}{\textbf{Baseline}} & \multicolumn{3}{c}{\textbf{\DL{}}} \\
\cmidrule(lr){2-4} \cmidrule(lr){5-7}
\textbf{Metric} & $N$ & $U$ & Quality & $N$ & $U$ & Quality \\
\midrule
Agency1         & $\rho=-0.041$ & $\rho=0.188$ & $\rho=0.079$ & $\rho=-0.116$ & $\rho=0.223$ & $\rho=0.085$ \\
Agency2         & $\rho=0.372^*$ & $\rho=0.408^{**}$ & $\rho=0.441^{**}$ & $\rho=0.231$ & $\rho=0.43^{**}$ & $\rho=0.406^{**}$ \\
\midrule
\end{tabular}
\end{table*}